\documentclass[journal]{IEEEtran}

\usepackage[T1]{fontenc}
\usepackage{soul}
\usepackage[utf8]{inputenc}
\usepackage{graphicx}
\usepackage{multicol}
\usepackage{multirow}
\usepackage{amsmath,amsfonts,amssymb}
\usepackage{mathtools}
\usepackage{gensymb}
\usepackage{graphicx}
\usepackage{wrapfig}
\usepackage{float}
\usepackage{algorithm}
\usepackage{algpseudocode}
\usepackage{enumerate}
\usepackage{url}
\usepackage{multirow}
\usepackage{wrapfig}
\usepackage{lipsum}
\usepackage{color}
\usepackage[colorlinks=true, allcolors=blue]{hyperref}
\usepackage{hhline} %
\usepackage{stfloats}
\usepackage{cite}

\ifCLASSINFOpdf
\else
\fi

\usepackage{xcolor}

\begin{document}

\title{
A forward model incorporating elevation-focused transducer properties for 3D full-waveform inversion in 
ultrasound computed tomography}
%
%
%

\author{Fu~Li,~\IEEEmembership{Student Member,~IEEE,}
        Umberto~Villa, 
        Nebojsa~Duric,\\ 
        
        and~Mark~A.~Anastasio,~\IEEEmembership{Senior Member,~IEEE}
\thanks{Manuscript received \today; revised ---.}
\thanks{
This work was supported in part by NIH under Award
R01EB028652, National Science Foundation’s Supercomputer Centers Program, the state of Illinois, and the University of Illinois.
}
\thanks{F. Li and M. Anastasio are with the Department
of Bioengineering, University of Illinois at Urbana-Champaign, Urbana,
IL, 61801 USA e-mail: \url{fuli2@illinois.edu, maa@illinois.edu}.}
\thanks{U. Villa is with the Oden Institute, University of Texas at Austin, Austin, TX, 78712 USA. e-mail: \url{uvilla@austin.utexas.edu}}
\thanks{N. Duric is with the Department of Imaging Sciences, University of Rochester, Rochester, NY, 14642, USA.}
}

\markboth{IEEE Transactions on Ultrasonics, Ferroelectrics, and Frequency Control
Submit Manuscript}%
{Fu \MakeLowercase{\textit{et al.}}: }

\maketitle

\begin{abstract}
Ultrasound computed tomography (USCT) is an emerging medical imaging modality that holds great promise for improving human health. Full-waveform inversion (FWI)-based image reconstruction methods account for the relevant wave physics to produce high spatial resolution images of the acoustic properties of the breast tissues.
A practical USCT design employs a circular ring-array comprised of elevation-focused ultrasonic transducers, and volumentric imaging is achieved  by translating the ring-array orthogonally to the imaging plane.
In commonly deployed slice-by-slice (SBS) reconstruction approaches, the three-dimensional (3D) volume is reconstructed by stacking together two-dimensional (2D) images reconstructed for each position of the ring-array.
A limitation of the SBS reconstruction approach is that it does not account for 3D wave propagation physics and the focusing properties of the transducers, which can result in significant image artifacts and  inaccuracies. To perform 3D image reconstruction when elevation-focused transducers are employed, a numerical description of the focusing properties of the transducers should be included in the forward model.
To address this, a 3D computational model of an elevation-focused transducer is developed to enable 3D FWI-based reconstruction methods to be deployed in ring-array-based USCT. The focusing is achieved by applying a spatially varying temporal delay to the ultrasound pulse (emitter mode) and recorded signal (receiver mode). The proposed numerical transducer model is quantitatively validated and employed in computer-simulation studies that demonstrate its use in image reconstruction for ring-array USCT.
\end{abstract}

\begin{IEEEkeywords}
Ultrasound computed tomography, lens-focused transducer,
3D full-waveform inversion
\end{IEEEkeywords}

\IEEEpeerreviewmaketitle

\vspace{-1mm}
\section{Introduction}
\label{sec:intro}
\IEEEPARstart{U}{ltrasound} computed tomography (USCT) is an emerging medical imaging technology that is being developed for a variety of clinical applications \cite{andre2013clinical,roy2013breast,guasch2020full,wiskin2020full,wiskin2022imaging,gierach2022rapid}.
For example, USCT methods for cancer detection and treatment response monitoring have been reported \cite{kratkiewicz2022ultrasound,littrup2021multicenter,duric2020novel,jw2017,wiskin2019quantitative,andre2013clinical,wiskin2022imaging}.
 Images that represent accurate estimates of the speed of sound (SOS), density, and acoustic attenuation (AA) of tissue can be produced by use of USCT image reconstruction methods ~\cite{pratt2007sound,wang2015waveform,matthews2017regularized, duric2021potential, duric2020using}.  
To account for the relevant wave physics and thereby achieve high spatial resolution images, full-waveform inversion (FWI) reconstruction methods\cite{wang2015waveform,bates2022probabilistic,lucka2021high,guasch2020full} are being developed.  Such advanced reconstruction methods can circumvent the limitations of simplified physics methods\cite{javaherian2021ray,hormati2010robust}. 

Recently, a variety of three-dimensional (3D) USCT measurement configurations have been proposed\cite{schreiman1984ultrasound,carson1981breast,andre1997high,duric2007detection,song2021design,stotzka2003ultrasound,GemmekeBergerHopp2018_1000079770,jw2017, malik2018quantitative, CUDEIROBLANCO20221995, 9512083,guasch2020full}. For example, a 3D USCT system utilizing a hemispherical array of transducers was developed by Ruiter \textit{et al.} 
\cite{stotzka2003ultrasound,GemmekeBergerHopp2018_1000079770}. A system developed by QT Ultrasound (QT Ultrasound, Inc, Novato, CA, USA)  employs a plane wave transmitter and a 2D flat detector \cite{jw2017, malik2018quantitative} to achieve 3D imaging.

Another practical USCT design that has been employed by multiple groups involves a circular ring-array containing elevation-focused transducer elements\cite{duric2007detection,duric2013breast,duric2014clinical,song2021design,8197331,9593866,mervcep2019transmission}. A few examples that employ this configuration include the SoftVue system (Delphinus Medical Technologies, Inc, Novi, MI, USA), approved by the Food and Drug Administration for the screening of women with dense breast tissue and diagnostic use for all women\cite{jcm11051165}, the UltraLucid system developed by Song \textit{et al.} \cite{song2021design,8197331} and a brain imaging system developed by Guasch \textit{et al.}\cite{CUDEIROBLANCO20221995,guasch2020full,9512083}.  The ring-array system enables acquisition of ultrasound measurements at multiple vertical positions 
by translating the ring of transducers vertically.
This design allows for slice-by-slice (SBS) image reconstruction, where a 3D volume is obtained by stacking together two-dimensional (2D) slices reconstructed from USCT measurements acquired at each vertical position of the ring-array \cite{matthews2017regularized,wang2015waveform,sandhu2015frequency}. 

While the SBS reconstruction approach for ring-array-based systems alleviates the computational burden associated with directly reconstructing a 3D volume, it possesses significant limitations.  
First and foremost, artifacts may appear in the reconstructed images due to scattering and diffraction from out-of-plane structures. Such effects are not accounted for in the 2D imaging model\cite{jw2017, sandhu2015frequency}, where point-like transducers are typically assumed and the spatial impulse responses (SIRs) of transducers are not modeled \cite{li2022investigation,poudel2019compensation}. 
Moreover, there  are also significant differences in the way acoustic waves propagate in 2D compared to 3D.
To reduce artifact levels and improve image quality in ring-array-based USCT, there remains an important need to develop reconstruction methods based on 3D wave physics models that can incorporate the elevational focusing properties of the transducers.

In this study, a 3D USCT forward model that accounts for the transducer's elevational focusing characteristics is developed to enable improved FWI-based image reconstruction in ring-array-based USCT.
In this way, the out-of-plane acoustic scattering and the focusing properties (i.e., the SIRs) of the transducers, in both transmit and receive modes, are accounted for in the forward and associated adjoint models that are utilized by image
reconstruction methods.
Specifically, a rigid flat transducer element coupled with a  concave acoustic lens is considered. The incorporation of focusing effects induced by an acoustic-lens in a numerical wavesolver-based USCT forward 
model is the main contribution of this work. This differs from related works \cite{martin2016simulating,wise2019representing} that are mostly concerned with accurate modeling of transducers with finite curved apertures that may not conform to the computational grid.

The proposed forward model is systematically validated by the use of an analytic solution. Additionally, the developed forward and adjoint models are utilized in a numerical case study of ring-array-based USCT that employ realistic 3D numerical breast phantoms (NBPs). 
This demonstrates, for the first time, the feasibility of incorporating transducer elevational focusing effects in a 3D time-domain FWI reconstruction method for ring-array-based USCT and reveals improvements in image quality over the traditional SBS reconstruction approach.
To enable related investigations by other researchers, the 3D NBPs and corresponding 3D USCT measurement data used in the case study have been made publicly available under CC-0 licensing \cite{li2021NBPs3D}.

The remainder of this article is organized as follows. In Section \ref{sec:background}, a brief review of the canonical USCT forward model and the time-domain FWI method is provided, as well as  background on ring-array-based USCT and  lens-focused transducer modeling. 
Section \ref{sec:method} introduces the proposed USCT forward model that incorporates elevational focusing effects. 
The forward model is validated and characterized as reported in Section \ref{sec:validation}.  
A case study is presented in Section \ref{sec:cases} to demonstrate the utility of the proposed forward and adjoint models for 3D FWI.
Section \ref{sec:dissusion} provides discussions of the numerical results presented in the case study, as well as of the limitations and future developments of the proposed approach.
Finally, a summary of the work are provided in Section \ref{sec:conclusion}.
\section{Background}
\label{sec:background}
\subsection{Canonical forward models for USCT}
\label{subsec2a}
USCT imaging models in their continuous-to-continuous (C-C) form and discrete-to-discrete (D-D) forms are reviewed below.
In USCT, a sequence of short acoustic pulses are transmitted from transducer emitters and subsequently
propagate through an object. 
The acoustic pulse is denoted as $x(t)\in \mathbb{L}^2([0, T])$ where $t \in [0, T]$ is the time coordinate, and $T$ is a fixed finite time interval. 
The spatiotemporal source $s_i(\textbf{r}, t) \in \mathbb{L}^2(\mathbb{R}^3 \times [0, T])$ that excites the $i$-th transducer is defined as
$s_i(\textbf{r}, t) = \mathcal{E}_i x(t) \equiv \chi_i(\textbf r) x(t)$ for $i = 0, 1, \ldots, M-1$.
Here, $\mathcal{E}_i: \mathbb{L}^2([0, T]) \mapsto \mathbb{L}^2(\mathbb{R}^3 \times [0, T])$ is the mapping operator, $\textbf{r}\in \mathbb{R}^3$ denotes the spatial coordinate, $\chi_i(\textbf r)$ is the indicator function describing the support of the active area of the $i$-th transducer surface, and $M$ is the total number of emitting transducers.

When the $i$-th source pulse $s_i(\textbf{r}, t)$ propagates through the object, it generates a pressure wavefield distribution denoted by $p_i(\textbf{r},t) \in \mathbb{H}^1(\mathbb{R}^3 \times [0, T])$\footnote{
Here, $\mathbb{H}^1(\mathbb{R}^3 \times [0, T])$ denotes the space of square integrable functions with square integrable spatiotemporal gradients. 
Regularity of weak solutions of the wave equation is studied e.g. in \cite[Ch 3]{lions2012non}. Specifically, a weak solution of the wave equation is such that $p \in \mathbb{L}^2(0,T; \mathbb{H}^1(\mathbb{R}^3))$ and $\frac{\partial p}{\partial t} \in \mathbb{L}^2(0,T; L^2(\mathbb{R}^3))$, where $\mathbb{L}^2(0,T; \mathbb{H}^1(\mathbb{R}^3)$ and $\mathbb{L}^2(0,T; \mathbb{L}^2(\mathbb{R}^3)$ are the Bochner spaces\cite{evans2012introduction}. It can be shown that if $p \in \mathbb{H}^1(\mathbb{R}^3 \times [0, T])$ then $p$ and $\frac{\partial p}{\partial t}$ belongs to the Bochner spaces above.}.
The wave propagation  in heterogeneous media can be described by the following lossy second-order wave equation \cite{treeby2010modeling}:
\begin{equation}
\vspace{-1mm}
\left\{
\begin{array}{rl}
&\frac{1}{c^2}\frac{\partial^2}{\partial t^2} p_i   - \rho\nabla \cdot \frac{1}{\rho} \nabla  p_i +  L \nabla^2 p_i = s_i \\
& L = \mu \frac{\partial}{\partial t} (-\nabla^2)^{\frac{y}{2}-1} + \eta(-\nabla^2)^{\frac{y-1}{2}}, \\
\end{array}
\right.\label{eqn:lossy_wave_eq}
\end{equation}
where $p_i = p_i(\textbf{r},t)$, $s_i = s_i(\textbf{r},t)$, and $c = c(\textbf{r})$, $\alpha = \alpha(\textbf{r})$, $\rho = \rho(\textbf{r})$ denote the heterogeneous  SOS, AA, and density distributions, respectively. The pseudo-differential operator $L$ models power-law frequency-dependent acoustic absorption and dispersion using fractional Laplacians. In particular, the first term stems from the Chen and Holm's reformulation \cite{chen2003modified} of Szabo's causal convolution operator \cite{szabo1995causal} and the second term is a dispersion correction derived from the Kramers–Kronig relations \cite{1503968}.
The quantities $\mu$ and $\eta$  denote absorption and dispersion proportionality coefficients,  that are defined as $\mu(\textbf{r}) =  -2\alpha(\textbf{r}) c(\textbf{r})^{y-1}$ and $ \eta(\textbf{r}) =  2\alpha(\textbf{r}) c(\textbf{r})^{y}\tan(\frac{\pi y}{2})$, where $y = y(\textbf{r})$ is power-law exponent.
Utilizing the fractional derivative enables the employment of non-integer power-law order $y$ and the fractional Laplacian operator enables efficient implementation through time-domain pseudospectral methods \cite{treeby2010modeling}. 

Equation \eqref{eqn:lossy_wave_eq} can be expressed in operator form as:
\begin{align}
\vspace{-1mm}
\label{eqn:forward_conti}
 p_i(\textbf{r},t) = \mathcal{H}^{\boldsymbol{u}(\textbf{r})} s_i(\textbf{r},t)\; 
\end{align}
where $s_i(\textbf{r},t) \coloneqq \mathcal{E}_i x(t)$.
The linear operator $\mathcal{H}^{\boldsymbol{u}(\textbf{r})}: \mathbb{L}^2(\mathbb{R}^3 \times [0, T]) \rightarrow \mathbb{H}^1(\mathbb{R}^3 \times [0, T])$ describes the action of the wave equation
and has an explicit dependence on the distribution of acoustic properties $\boldsymbol{u}(\textbf{r}) =[c(\textbf{r}), \alpha(\textbf{r}), \rho(\textbf{r}), y(\textbf{r})]$.
%

During data acquisition, $p_i (\textbf{r},t)$ is only recorded at a limited number of transducer receivers. These measurement data will be referred to as $g_{ij}(t) \in L^2([0, T])$, with the subscript $i$ denoting that the data were produced by the $i$-th source, also referred to as the $i$-th data acquisition, and the subscript $j$ denoting the data recorded at the $j$-th transducer for $j = 0, 1, \ldots, M-1$.  In the case where discrete sampling effects and transducer focusing properties are not considered, the idealized C-C forward model can therefore be described as:
\begin{align}
g_{ij}(t) =  \mathcal{M}_j{\mathcal{H}}^{\boldsymbol{u}(\textbf{r})}s_i(\textbf{r},t) \; 
 \text{ with }  s_i(\textbf{r},t) = \mathcal{E}_i x(t) ,
\label{eqn:forward_gHf}
\end{align}
for all emitters $i=0,1, \ldots, M-1$ and all receivers $j=0,1, \ldots, M-1$.
Here, $\mathcal{M}_j: \mathbb{H}^1(\mathbb{R}^3 \times [0, T]) \mapsto L^2([0, T])$ extracts the measurements corresponding to the
$j$-th transducer location on the measurement aperture; specifically, $g_{ij}(t) = \mathcal{M}_j p_i(\textbf{r}, t) \equiv \int_{\mathbb{R}^3} \chi_j(\textbf{r}) p_i(\textbf{r}, t) \, d\,\textbf{r}$. Note that, because of acoustic reciprocity, the mapping operator $\mathcal{E}_i$ of the $i$-th transducer is the adjoint of the sampling operator $\mathcal{M}_i$, i.e. $\mathcal{E}_i = \mathcal{M}_i^\dag$.

The description of a digital imaging system is typically approximated in practice by a D-D imaging model. To establish this, $\boldsymbol{u}(\textbf{r})$, $x(t)$, $s_i(\textbf{r}, t)$, $p_i(\textbf{r},t)$, $g_{ij}(t)$ are sampled on a Cartesian grid and at a temporal interval $\triangle t$ to obtain the finite-dimensional representations as $\textbf{u} \in \mathbb{R}^{4K}$, $\textbf{x} \in \mathbb{R}^{L}$, $\textbf{s}_i \in \mathbb{R}^{KL}$, $\textbf{p}_i \in \mathbb{R}^{KL}$, and $\textbf{g}_{ij} \in \mathbb{R}^{L}$.
Here, $K$ and $L$ denote the number of spatial and temporal samples, respectively. 
Let $[\mathbf{z}]_k$ denote the $k$-th element of a vector $\mathbf{z}$.
These discretized parameters can be defined as:
\begin{align}
    \nonumber
  &[\textbf{u}]_{k} \coloneqq c(\textbf{r}_k),  \,
  [\textbf{u}]_{K+k} \coloneqq \alpha(\textbf{r}_k), \,
  [\textbf{u}]_{2K+k} \coloneqq \rho(\textbf{r}_k),\\
   \nonumber
  &[\textbf{u}]_{3K+k} \coloneqq y(\textbf{r}_k),\,
  [\textbf{x}]_l \coloneqq x(t_l),\,
  [\textbf{s}_i]_{kL+l} \coloneqq s_i(\textbf{r}_k, t_l),\\
   \nonumber
  &[\textbf{p}_i]_{kL+l} \coloneqq p_i(\textbf{r}_k, t_l), \quad \text{and} \quad [\textbf{g}_{ij}]_l \coloneqq  g_{ij}(t_l),
\end{align}
for $k = 0,1,\ldots, K-1$, $ l = 0,1, \ldots, L-1$,
where $\textbf{r}_k$ denotes the  $k$-th spatial grid point and $t_l = l\, \triangle t$ denotes the $l$-th time sample.
Given these discretized quantities, a D-D version of the idealized imaging model in
\eqref{eqn:forward_gHf} can be expressed as:  
\begin{align}
\small
\label{eqn:forward_DD}
\textbf{g}_{ij} =  \textbf{M}_j{\textbf{H}}^\textbf{u} \textbf{s}_i \; \text{ with } \textbf{s}_i = \textbf{M}_i^\intercal\textbf{x},
\end{align}
for all emitters $i=0, 1, \ldots, M-1$ and all receivers $j=0, 1, \ldots, M-1$.
Above, the matrix ${\textbf{H}}^\textbf{u} \in \mathbb{R}^{KL \times KL}$ represents a discrete approximation of the wave propagation operator ${\mathcal{H}}^{\textbf{u}(\textbf{r})}$, whose action is implemented by use of a numerical wave solver method. The matrix $\textbf{M}_j \in \mathbb{R}^{L \times  KL}$ denotes the discretization of the sampling operators $\mathcal{M}_j$ corresponding to the $j$-th transducer, and the superscript $\cdot^\intercal$ denotes the transpose operation. Specifically, $\textbf{M}_j= \tilde{\textbf{M}}_j \otimes \textbf{I}_L$, where $\tilde{\textbf{M}}_j \in \mathbb{R}^{K \times 1}$ stems from the discretization of the indicator function $\chi_j(\textbf{r})$ (i.e. $[\tilde{\textbf{M}}_j]_{k} = \chi_j(\textbf{r}_k)$ for $k=0,1, \ldots, K-1$), $\textbf{I}_L \in \mathbb{R}^{L\times L}$ is the $L$-dimensional identity matrix, and $\otimes$ denotes the Kronecker product. In the special case where the $j$-th transducer is point-like and located at the position $\textbf{r}_k$, the entries of the matrix $\tilde{\textbf{M}}_j$ are all zero but the $k$-row entry that is equal to 1.

Finally, by collecting in a single vector $\textbf{g}_{i} \in \mathbb{R}^{ML}$ the measurement data $\textbf{g}_{ij}$ recorded by all receivers when the $i$-th emitter is excited, the discrete imaging model in \eqref{eqn:forward_DD} can be rewritten in a more compact form as 
\begin{equation}
\begin{split}
\small
\label{eqn:forward_DD2}
\textbf{g}_{i} =  \textbf{M}{\textbf{H}}^\textbf{u} \textbf{s}_i, 
\end{split}
\end{equation}
where $\textbf{M} \in \mathbb{R}^{ML \times KL}$ is defined as $ [\textbf{M}_0; \textbf{M}_1;  \ldots; \textbf{M}_{M-1}]$, and $\textbf{g}_i \in \mathbb{R}^{ML}$ is defined as $\textbf{g}_i = [ \textbf{g}_{i0}; \textbf{g}_{i1}; \ldots; \textbf{g}_{i(M-1)} ]$. Here, the semicolon notation denotes row-wise concatenation. 



\subsection{Time-domain waveform inversion with source encoding (WISE) in its discrete form} 
\label{sec:wise}
The USCT reconstruction problem is to estimate the tissue acoustic properties $\textbf{u}$, or a subset of them, from a collection of (noisy) measurement data $\underline{\textbf{g}_i} \in \mathbb{R}^{ML}$, for $i = 0, 1, \ldots, M-1$.  The underlined notation denotes that $\underline{\textbf{g}_i}$ is measured data as opposed to the simulated data ${\textbf{g}_i}$ produced by use of the forward model.
This problem can be solved by use of the WISE method \cite{wang2015waveform,zhang2012efficient,moghaddam2013new}. The WISE method is a FWI method that circumvents the large computational burden of conventional FWI \cite{wang2015waveform,lucka2021high} by leveraging the superposition of acoustic waves corresponding to the excitation of multiple emitters. 
When an $\ell^2$-norm is employed as the misfit functional, the WISE method can be formulated as a stochastic optimization problem as \cite{wang2015waveform}:
\begin{equation}
\begin{split}
\hat{\textbf{u}} 
&= \arg \min_\textbf{u}  \mathbb{E}_\textbf{w}{\frac{1}{2}||\underline{\textbf{g}^\textbf{w}}-\textbf{M}{\textbf{H}}^\textbf{u}\textbf{s}^\textbf{w}  ||_2^2} + \lambda R(\textbf{u}),
\end{split}
\label{eqn:WISE}
\end{equation}
where $\textbf{w} \in \mathbb{R}^{M}$ is a random encoding vector with zero mean and identity covariance matrix, $\mathbb{E}_\textbf{w}$ denotes the expectation w.r.t $\textbf{w}$, $\lambda$ is a regularization parameter and $R(\textbf{u})$ is a regularization penalty. The quantities
$
\underline{\textbf{g}^\textbf{w}} = \sum_{i=0}^{M-1}[\textbf{w}]_i \underline{\textbf{g}_i}$ and $ \textbf{s}^\textbf{w} = \sum_{i=0}^{M-1}[\textbf{w}]_i \textbf{s}_i = \sum_{i=0}^{M-1}[\textbf{w}]_i \textbf{M}^\intercal_i\textbf{x}
$
are the encoded measurement data and the encoded source, respectively. In previous studies, the encoding vector $\textbf{w}$ has been chosen according to a Rademacher distribution \cite{wang2015waveform,krebs2009fast}.

\subsection{Modeling of lens-focused transducers based on geometrical acoustics approximation}
\label{sec:lens-bg}

In ring-array USCT systems, an elevational focusing effect can be achieved by use of an acoustic lens attached to the front face of a flat transducer element \cite{smith1978real,song2013liquid,duric2014clinical}.
The variation in the thickness of a concave lens introduces a spatially dependent time-delay in the wavefield upon propagation through the lens. 

One approach to modeling a flat transducer is based on the Rayleigh–Sommerfeld integral \cite{o1949theory,hansen2001fundamentals}.
In transmit mode, the generated acoustic pressure can be described as the superimposition of the contributions of point sources that span the transducer aperture $\mathbf{\Omega} \subset \mathbb{R}^3 $, which describes the active area of the transducer element.
An acoustic source on an aperture may be modeled as a velocity source or a pressure source.
When an aperture surface $\mathbf{\Omega}$ of an otherwise perfectly rigid boundary vibrates with a normal velocity $v_{\bot}(\textbf{r},t)$,
the acoustic pressure $p(\textbf{r}, t)$ radiated from $\mathbf{\Omega}$, without a lens present, into a homogeneous non-lossy medium with constant SOS $c_0$ and density $\rho_0$   can be analytically described as:
\begin{equation}
\resizebox{.89\hsize}{!}{$
 p(\textbf{r}, t) = \rho_0  \iint_\mathbf{\Omega}{ 
\frac{1}{2\pi ||\textbf{r}-\textbf{r}'||} \frac{\partial}{\partial t} v_{\bot } \left(\textbf{r}', t-\frac{||\textbf{r}-\textbf{r}'||}{c_0}\right) d\textbf{r}'}$}.
\label{eqn:solution1}
\end{equation}

Equation (\ref{eqn:solution1}) can be generalized to the case where a lens is attached to the transducer by incorporating a time delay into the normal velocity term.
One way to accomplish this is to invoke a geometrical acoustics approximation to describe the interaction of the pressure wavefield with the lens, and neglect any effects associated with mode conversion into shear waves and refraction at the lens surface\cite{Guyomarfocu,Penttinen_1976,4154642,wu2002theoretical,911727, marechal2007lens}. The impacts associated with attenuation of the lens and impedance mismatches between the lens and propagation medium can be modeled by introducing an apodization weights function, which is discussed in Appendix \ref{app:impedence} but omitted in this section.

In this case, the transducer properties introduced by the curvature of the lens can be modeled implicitly by defining the normal component of the velocity as
\begin{equation}
    v_{\bot } \left(\textbf{r}', t \right) = \hat{v}\left(t+\tau(\textbf{r}') \right)  \quad \forall \textbf{r}' \in \mathbf{\Omega}.
    \label{eq:lens_pvel} 
\end{equation}
Here, the quantity $\hat{v}(t)$ denotes the normal velocity of the transducer element front face that is constant in space over the aperture $\mathbf{\Omega}$, and $\tau(\textbf{r}')$ is a spatially varying time shift. 
%
%
The time shift is determined by the  thickness profile $d(\textbf{r}')$  and SOS, $c_{lens}$, of the concave lens as \cite{4154642,wu2002theoretical,911727}:
\begin{align}
\label{eqn:timedelay}
    \tau(\textbf{r}') = d(\textbf{r}')\left(\frac{1}{c_{0}}-\frac{1}{c_{lens}}\right). 
\end{align}

In receive mode, the recorded signal is affected by
  the finite detecting aperture, again referred to as $\mathbf{\Omega}$, and the time-delays introduced by the lens mounted on it \cite{kostli2003two}.  
In this case, the measured signal $g_{\mathbf{\Omega}}(t)$ can be analytically described by an integral of the incident pressure wavefield evaluated over  $\mathbf{\Omega}$ \cite {ding2017efficient,5560859} with time delays defined in \eqref{eqn:timedelay}:
\begin{equation}
\begin{aligned}
 g_{\mathbf{\Omega}}(t) &= \iint_\mathbf{\Omega} \,p\left(\textbf{r},t+\tau(\textbf{r}) \right) d \textbf{r}.
\label{eqn:c_receiver} 
\end{aligned}
\end{equation}


In practical applications of USCT, the acoustic medium is spatially heterogeneous and analytic expressions for $p(\textbf{r}, t)$ are generally not available. Simulation of the transmitted or received measurement signal therefore requires use of a numerical wavesolver method as described below.
 
\section{USCT forward model for use with elevation-focused transducers}
\label{sec:method}
In this section, a new forward model for ring-array USCT that incorporates the elevational focusing properties of the transducers is formulated.
The forward model is presented first in a C-C form then in its D-D form. The adjoint model for use in FWI is also provided in a D-D form.

\subsection{The C-C forward model incorporating focusing effects}
Here, a lens-focused transducer model will be combined with a wave propagation model to
establish an overall forward model for lens-focused USCT. 
To accomplish this, the relationship between the source pulse $s_i(\textbf{r}, t)$ in \eqref{eqn:lossy_wave_eq} and the normal velocity $v_{\bot}(\textbf{r},t)$ in the transducer model \eqref{eqn:solution1} is described and a new sampling operator and its adjoint are introduced.

For the rigid baffle transducer model employed in Section \ref{sec:background}, the forcing term $s_i(\textbf{r},t)$ can be modeled as a mass source in \eqref{eqn:lossy_wave_eq} \cite{verweij2014simulation}.
Specifically, the forcing term has the form \begin{equation}
s_i(\textbf{r},t) = \chi_i(\textbf r)\frac{\partial}{\partial t}S_m (\textbf{r},t) = 2\hat{\rho_0}\chi_i(\textbf r) \frac{\partial}{\partial t} v_{\bot}(\textbf{r},t),
\label{eq:mass_source}
\end{equation}
where $\hat{\rho_0}$ is the ambient density, $S_m(\textbf{r},t) = 2\hat{\rho_0}v_{\bot}(\textbf{r},t)$ is a mass source, 
$v_{\bot}(\textbf{r},t)$ 
and $\chi_i(\textbf{r})$  are the normal component of the velocity and the indicator function describing the support of the $i$-th transducer as introduced above \cite{verweij2014simulation,treeby2010k}. 

By use of \eqref{eq:lens_pvel},  \eqref{eq:mass_source} can be re-expressed as:
\begin{equation}
s_i(\textbf{r}, t) =  2\textbf{$\hat{\rho_0}$}\chi_i(\textbf r)\frac{\partial}{\partial t} \hat{v}(t + \tau_i(\textbf{r})),
\label{eq:mass_source2}
\end{equation}
where $\tau_i(\textbf{r})$ is the time delay function associated with the $i$-th transducer. Equation \eqref{eq:mass_source2} defines the excitation source in the lens-focused USCT forward model provided below.

This source can be further characterized in terms of a  mapping operator $(\mathcal{M}^{\tau}_i)^\dag : \mathbb{L}^2([0, T]) \mapsto \mathbb{L}^2(\mathbb{R}^3 \times [0, T])$. First, 
let $D_{\tau}: \mathbb{L}^2([0, T]) \mapsto \mathbb{L}^2([0, T])$ denote a time shift operator that satisfies  $\phi_\tau(t) = D_{\tau}\phi(t) \coloneqq  \phi(t+\tau)$, where $\phi(t)$  is a one-dimensional (1D) continuous signal.
Additionally, 
let $x(t) \coloneqq 2\,\hat{\rho_0}\frac{\partial}{\partial t}\hat{v}(t)$.
In terms of these quantities, the source and operator $(\mathcal{M}^{\tau}_i)^\dag$ are related as:
\begin{equation}
\label{contin_mapping_M} 
    s_i(\textbf{r},t) = (\mathcal{M}^{\tau}_i)^\dag x(t) \coloneqq \chi_i(\textbf r)D_{\tau_i(\textbf{r})}x(t).
\end{equation}

The measurement data $g_{ij}(t)$ produced by the $i$-th source
and measured by the $j$-th transducer can be  
described by use of the
 sampling operator $(\mathcal{M}^{\tau}_j)$ as:
\begin{equation}
\label{contin_sampling_M}
    g_{ij}(t) = (\mathcal{M}^{\tau}_j)  p_i(\textbf{r}, t) \coloneqq \int_{\mathbb{R}^3} \chi_j(\textbf{r})  D_{\tau_j(\textbf{r})}p_i(\textbf{r}, t) d\textbf{r}.
\end{equation}

Finally, the idealized C-C forward model in  \eqref{eqn:forward_gHf}  
can be generalized to the case of elevational-focused USCT as:
\begin{align}
\label{contin_gHs}
\resizebox{.89\hsize}{!}{$
g_{ij}(t) =  \mathcal{M}^\tau_j{\mathcal{H}}^{\boldsymbol{u}(\textbf{r})}s_i(\textbf{r},t) \text{ with } s_i(\textbf{r},t) = (\mathcal{M}^{\tau}_i)^\dag x(t).
$}
\end{align}

\subsection{The D-D imaging model incorporating focusing effects}
Here, a D-D version of   \eqref{contin_gHs}  is provided.
First, the discrete counterpart of the time delay operator $D_{\tau}$, denoted as $\textbf{D}_{\mathbf{\tau}} \in \mathbb{R}^{L \times L}$,
is defined as:
\begin{equation}
\textbf{D}_{\mathbf{\tau}} \boldsymbol {\phi} \coloneqq  \textbf{F}^{-1} \left[ exp(-j2\pi \textbf{k} \mathbf{\tau}) \odot (\textbf{F}\boldsymbol{\phi}) \right],
\label{eq:discreteTimeDelayOp}
\end{equation}
where $\boldsymbol{\phi} \in \mathbb{R}^L$ denotes a 1D discrete temporal signal, $\textbf{F} $ and $\textbf{F}^{-1}$ denote the 1D discrete Fourier transform and its inverse,
$\odot$ represents element-wise multiplication, and $\textbf{k}$ is a set temporal frequencies defined as:
\begin{align}
\nonumber
\textbf{k} = \left\{\begin{array}{ll} 
\frac{2\pi}{L \Delta t}\left[ -\frac{L-1}{2}, -\frac{L-1}{2}+1, ..., \frac{L-1}{2} \right]   \, &{\rm if} \, L \, \text{is odd} 
\\[3pt] 
\frac{2\pi}{L \Delta t}\left[ -\frac{L}{2}, -\frac{L}{2}+1, ..., \frac{L}{2} \right]   \,  &{\rm if} \, L \, \text{is even.}\end{array}
\right.
\end{align}
As defined in Section  \ref{subsec2a}, $L$ denotes the number of
temporal samples of the signal that are spaced by $\Delta t$.
Note that \eqref{eq:discreteTimeDelayOp}  allows for  arbitrary time delays to be implemented that are not necessarily exact multiples of the sampling interval $\Delta t$.

\begin{figure}[htb]
\centering
\includegraphics[width=0.45\textwidth]{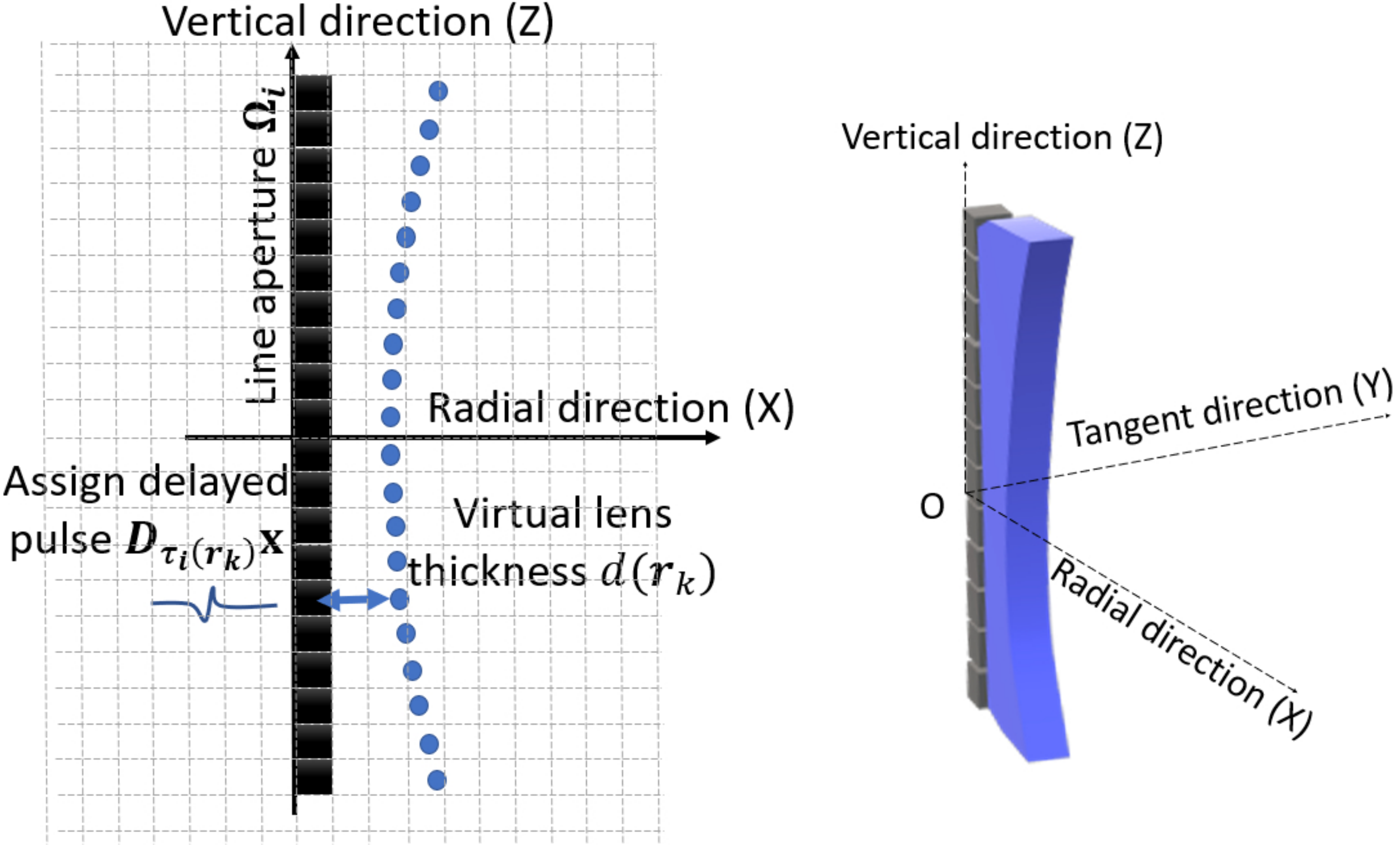}
\caption{Schematic of a discretized focused transducer. The transducer aperture
is approximately a line segment due to the high aspect
ratio. The vertical, radial and tangent directions are referred as the Z-, X- and Y-axis, respectively. The radial direction is toward the center of the imaging system.}
\label{fig:trans}
\end{figure}

Next, the transducer aperture $ \mathbf{\Omega}$ is discretized to conform to the assumed computational grid. Due to the high aspect ratio of the transducers often employed in ring-array USCT \cite{duric2014clinical,duric2013breast}, the transducer width is neglected and the aperture is described approximately by a line that is parallel to the vertical axis and whose length corresponds to the height of the transducer.
The line-aperture is divided into $N$ consecutive line-segments, with the length of each segment corresponding to height of a voxel in the computational grid. An example of a lens-focused transducer and such segmented line-aperture is depicted in Fig. \ref{fig:trans}. When the line-aperture does not evenly bisect the voxels, a nearest-neighbor interpolation method is employed where each segment is assigned to the nearest grid point.
Hereafter, the set $\mathcal{K}_{\mathbf{\Omega}_i} \subset \{0, 1,\ldots, K-1\}$ will denote the collection of indexes $k$ corresponding to the grid points associated with the $i$-th transducer aperture ($\textbf{r}_k \in \mathbf{\Omega}_i$). The cardinality of $\mathcal{K}_ {\mathbf{\Omega}_i}$ is the number $N$ of consecutive line-segments that form the line-aperture.

Next, as a prerequisite for establishing the D-D forward model, discrete versions of \eqref{contin_mapping_M} and \eqref{contin_sampling_M} are established. Firstly, a sampling matrix $\textbf{M}_j^{\boldsymbol{\tau}}$ corresponding to the $j$-th receiver is introduced as the discrete counterpart of $\mathcal{M}^{\tau}_j$.
It is defined as  $\textbf{M}_j^{\boldsymbol{\tau}} \coloneqq \sum_{k \in \mathcal{K}_{\mathbf{\Omega}_j} } \tilde{\textbf{M}}^k \otimes \textbf{D}_{{\tau}_j(\textbf{r}_k)}$, where
$\tau_j(\textbf{r}_k)$ is the time delay of the $j$-th transducer at a position $\textbf{r}_k$, and  $\tilde{\textbf{M}}^k \in \mathbb{R}^{K \times 1}$ is an indicator vector.
Specifically, $[\tilde{\textbf{M}}^k]_i = \delta_{ik}$ such that the $k$-th element is equal to 1 and other elements are all zero, where $\delta_{ik}$ is a Kronecker delta function.
Similarly, the mapping matrix for the $i$-th transmitter, which is the transpose of the sampling matrix,  
is defined as   $(\textbf{M}_i^{\boldsymbol{\tau}})^ \intercal \coloneqq  \sum_{k \in \mathcal{K}_{\mathbf{\Omega}_i} } ({\tilde{\textbf{M}}^k})^\intercal \otimes \textbf{D}_{{\tau}_i(\textbf{r}_k)}$. Note that the time shift operator is self-adjoint and the Kronecker product is commutative.


To establish the discrete counterpart of \eqref{contin_mapping_M}, the spatial-temporal source term  
is specified by use of the mapping matrix $(\textbf{M}_i^{\boldsymbol{\tau}})^ \intercal$. Specifically, for the $i$-th transmitter, 
the source term is defined as: 
\begin{equation}
\resizebox{.89\hsize}{!}{$
\begin{split}
 \begin{aligned}
\label{eqn:HsirE}
[\textbf{s}_i]_{kL+l} = [(\textbf{M}_i^{\boldsymbol{\tau}})^\intercal \textbf{x}]_{kL+l} \coloneqq &\left\{
 \begin{array}{ll}
[\textbf{D}_{{\tau}_i(\textbf{r}_k)}\textbf{x}]_l   & \forall k \in \mathcal{K}_{\mathbf{\Omega}_i}  \\
 0 &  \forall k \notin \mathcal{K}_{\mathbf{\Omega}_i},
 \end{array}\right.  
\end{aligned}
\end{split}
$}
\end{equation}
for
$    k = 0, 1, \ldots, K-1 $ and
$    l = 0, 1,  \ldots, L-1
$,
where $\textbf{x} \in \mathbb{R}^L$ is a 1D discretized signal temporally sampled from the continuous velocity source as $[\textbf{x}]_l \coloneqq 2 \hat{\rho}_0\frac{\partial }{ \partial t}\hat{v}(t_l)$. 
In  \eqref{eqn:HsirE}, the source $\textbf{s}_i$ can be interpreted as a superposition of temporally-delayed contributions produced by the $N$ line-segments belonging to the $i$-th transducer $\mathbf{\Omega}_i$.

To establish the discrete counterpart of \eqref{contin_sampling_M}, 
the pressure data recorded by receivers is specified by use of the sampling matrix $\textbf{M}^{\boldsymbol{\tau}}_j$. 
Specifically,
the discrete pressure data $\textbf{g}_{ij}$ that are produced by the $i$-th source and recorded at the $j$-th transducer are described as
\begin{align}
\label{DD_receiver}
\textbf{g}_{ij} = \textbf{M}^{\boldsymbol{\tau}}_j \textbf{p}_i \coloneqq \sum_{k \in \mathcal{K}_{\mathbf{\Omega}_j}} \textbf{D}_{\tau_j(\textbf{r}_k)}\textbf{p}_i(\textbf{r}_k),
\end{align}
where $\textbf{p}_i(\textbf{r}_k)$ denotes 1D temporal discrete pressure data at the grid point $\textbf{r}_k$ generated by source $\textbf{s}_i$,  and the summation is over the $j$-th receiver aperture $\mathbf{{\Omega}}_j$.
In   \eqref{DD_receiver}, the received data are formed as a superposition of discrete pressure signals corresponding to different locations on the transducer aperture, each with appropriately defined time shifts.

Finally, a D-D version of the C-C forward model in \eqref{contin_gHs}
is obtained as
\begin{align}
 \textbf{g}_i = \textbf{M}^{\boldsymbol{\tau}} \textbf{H}^{\textbf{u}} \textbf{s}_i
\text{ with }
 \textbf{s}_i = (\textbf{M}_i^{\boldsymbol{\tau}})^\intercal \textbf{x},
 \label{eqn:new_DD_forward}
\end{align}
where $\textbf{M}^{\boldsymbol{\tau}} \coloneqq [\textbf{M}^{\boldsymbol{\tau}_0}_0; \textbf{M}^{\boldsymbol{\tau}_1}_1;  \ldots; \textbf{M}^{\boldsymbol{\tau}_{M-1}}_{M-1}]\in \mathbb{R}^{ML \times KL}$.
This equation represents a generalization of the idealized D-D forward model in  \eqref{eqn:forward_DD2} to the case of elevational-focused USCT.

\subsection{Formulation of the adjoint equation} 

Similar to the formulation reviewed in Section  \ref{sec:wise} and reusing the same notation,
the reconstruction problem in 3D ring-array based USCT can be defined as:
\begin{equation}
\begin{split}
\hat{\textbf{u}} 
&= \arg \min_\textbf{u}  \mathbb{E}_\textbf{w}{\frac{1}{2}||\underline{\textbf{g}^\textbf{w}}-\textbf{M}^{\boldsymbol{\tau}}{\textbf{H}}^\textbf{u}\textbf{s}^\textbf{w}  ||_2^2} + \lambda R(\textbf{u}).
\end{split}
\label{eqn:WISE_MM}
\end{equation}

To enable reconstruction of the SOS distribution, the gradient of data fidelity term in   \eqref{eqn:WISE_MM} with respect to $\textbf{c}$, denoted by $\nabla_\textbf{c} \textbf{J}^\textbf{w}$, can be computed via an adjoint state method \cite{norton1999iterative, plessix2006review}. The discretized expression for the gradient is given by \cite{perez2017time, wang2015waveform, lucka2021high}:
\begin{equation}
\resizebox{.88\hsize}{!}{$
\begin{split}
\label{eqn:sos_gradient}
[\nabla_\textbf{c} \textbf{J}^\textbf{w}]_k = &\frac{1}{[\textbf{c}]^3_k} 
\sum^{L-2}_{l=1} [\textbf{q}^\textbf{w}]_{kL+(L-l)}\\
&\times\frac{\left ([\textbf{p}^\textbf{w}]_{kL+l-1}-2[\textbf{p}^\textbf{w}]_{kL+l}+[\textbf{p}^\textbf{w}]_{kL+l+1}\right )}{\triangle t},
\end{split}
$}
\end{equation}
where $k$, $l$ are the indices of grid points and time steps, respectively.  
The quantity $\textbf{p}^\textbf{w} \in \mathbb{R}^{KL}$ denotes the wavefield data generated by the encoded source as $\textbf{p}^\textbf{w} = \textbf{H}^{\textbf{u}} \textbf{s}^\textbf{w}$ 
and   $\textbf{q}^\textbf{w}  \in \mathbb{R}^{KL}$ denotes the adjoint wavefield data that are computed as:
\begin{align}
\begin{split}
    \textbf{q}^\textbf{w} &= \textbf{H}^{\textbf{u}} (\textbf{M}^{\boldsymbol{\tau}})^\intercal \boldsymbol{\eta}^\textbf{w};  \\
    \text{ with } [\boldsymbol{\eta}^\textbf{w}]_{jL+l} & \coloneqq [\textbf{M}^{\boldsymbol{\tau}}{\textbf{H}}^\textbf{u}\textbf{s}^\textbf{w} - \underline{\textbf{g}^\textbf{w}}]_{jL+(L-l)}.
    \end{split}
\end{align}
Here, $\boldsymbol{\eta}^\textbf{w} \in \mathbb{R}^{ML}$ is the time-reversed data residual and $j$ is the index of the transducer.

Equation \eqref{eqn:sos_gradient} holds when non-lossy media is considered. 
In attenuating media,  \eqref{eqn:sos_gradient} is an approximate expression for the gradient with respect to the SOS,  as additional terms coming from AA modeling are neglected \cite{perez2017time}.  

\section{Validation and characterization of the imaging models}
\label{sec:validation}
In this section, the proposed forward model is validated and characterized. 
First, the forward model is validated by use of a semi-analytical approximation of the Rayleigh-Sommerfield diffraction integral solution for the case of a lossless homogeneous medium. 
Next, a sensitivity map is computed to describe the region of a 3D object that is effectively ``seen'' by a specified focused emitter-receiver pair in a ring-array system. 

\begin{figure}[htb]
    \centering
    \includegraphics[width=0.39\textwidth]{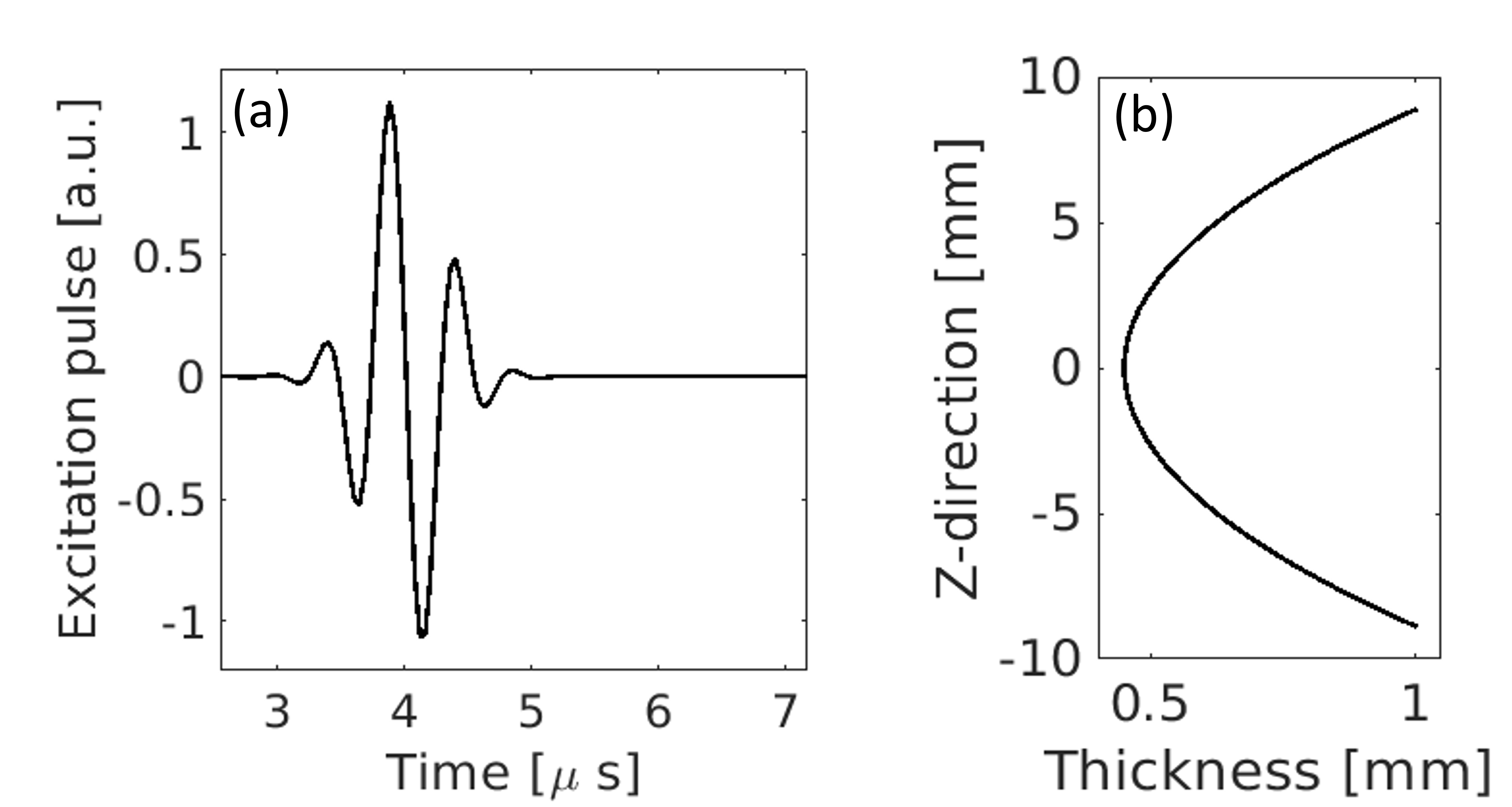}
        \vspace{-4mm}
    \caption{(a) The excitation source pulse employed in the validation studies. (b) The profile of the lens thickness $d(z) = \frac{a}{2}z^2 +d_0$ with curvature $a = 0.014 \text{ mm}^{-1}$ and offset $d_0 = 0.45 \text{ mm}$. Note that plot is not to scale and the horizontal axis representing the thickness of the lens has been stretched for visualization purposes.}
    \label{fig:source}
\end{figure}

In the following studies, all emitters and receivers were specified using the same elevation-focused transducer model. Specifically, a line transducer with a height of 18 mm and a  concave lens with a parabolic curvature was considered. The SOS in the lens was set as $c_{lens} = 4500 \text{ m/s}$. The shape of the excitation source is shown in Fig. \ref{fig:source}(a), and the profile of the lens is shown in Fig. \ref{fig:source}(b). 
The source pulse had a central frequency of $2 \text{ MHz}$ and a maximum frequency of $3.5  \text{ MHz}$.
Additional information regarding the employed source pulse, acoustic lens curvature, and associated time delays can be found elsewhere \cite{li2022investigation}.

\subsection{Validation of the forward model}


In the first validation study, the pressure wavefield produced by use of the numerical transducer model and measured by use of an ideal point transducer was compared to  the reference solution described in \eqref{eqn:solution1}-\eqref{eqn:timedelay}.
This is equivalent to validating the source $\textbf{s}_i = (\textbf{M}_i^{\boldsymbol{\tau}})^\intercal \textbf{x}$ and operator $\textbf{H}^{\textbf{u}} \textbf{s}_i$.
The reference solution was computed in a semi-analytical way, where the integrals were computed numerically by the distributed point source method\cite{5610565}.
\begin{figure*}[htb]
\centering
\includegraphics[width=1\textwidth]{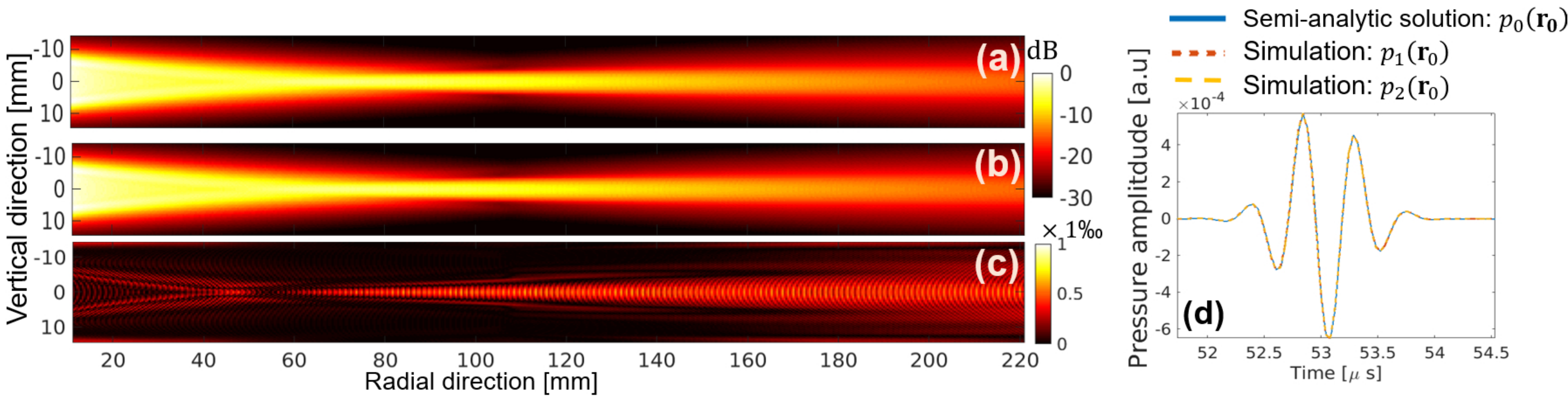}
\vspace{-4.6mm}
\caption{The maximum pressure amplitude maps in a transverse plane are displayed, using (a) the numerical transducer model and (b) the semi-analytical reference solution. These maps are presented on a decibel scale and computed using the formula $20\log_{10} { \frac{\max_{t} {|p_i|}}{\max_{\textbf{r},t} {|p_i|}} }$ ($i=1,2$), where $p_0 = p_0(\textbf{r},t)$ and $p_1 = p_1(\textbf{r},t)$ represent the pressure fields computed by the semi-analytic solution and the proposed method, respectively.
The relative difference map of the two solutions is shown in (c). The map is normalized by the maximum value of the on-axis (z = 0 mm) semi-analytic solution ($\frac{|\max_t p_0-\max_t p_1|}{|\max_t p_0(\textbf{r}_1,t)|}$), where
$\textbf{r}_1 = (x, 0, 0)$).
The focal region is situated at a distance ranging from 60 to 140 mm from the transducer.
(d) Validation of the operator $\textbf{H}^\textbf{u} \textbf{s}_i$ ($p_0(\textbf{r}_0)$ vs $p_1(\textbf{r}_0)$) and the overall forward operator $\textbf{M}_j^{\boldsymbol{\tau}} \textbf{H}^{\textbf{u}} \textbf{s}_i$ is conducted using acoustic reciprocity ($p_0(\textbf{r}_0)$ vs $p_2(\textbf{r}_0)$). Here, $p_0(\textbf{r}_0)$ and $p_1(\textbf{r}_0)$ denote the pressure measured at a specific location $\textbf{r}_0$ (x = 71 mm, y = 1 mm, z = 0.1 mm), while $p_2(\textbf{r}_0)$ represents the response of the lens-focused transducer when a point-like source is at location $\textbf{r}_0$. The source pulse depicted in Fig. \ref{fig:source}(a) was employed to compute all pressure fields $p_0$, $p_1$, and $p_2$.}

\label{fig:transducer_compare}
\end{figure*}
The action of the operator $\textbf{H}^{\textbf{u}}$
was implemented  by use the k-space pseudospectral method\cite{treeby2012modeling,treeby2010k}. 
The computational domain was a uniform 3D Cartesian grid with $1280 \times 1280 \times 144$ cubic voxels. Each voxel had a dimension of $0.2 \text{ mm}$. Accordingly, the transducer was divided into a 1D array of 90 voxels, corresponding to a height of 18 mm. 
The discretized source term $\textbf{s}_i$ with varying time delays was assigned to the 90 voxels as multiple mass sources.
To ensure accurate simulation using pseudospectral methods, the spatial discretization of the pressure wavefield was set to 3 points per wavelength \cite{treeby2012modeling}.
To ensure numerical stability, the Courant–Friedrichs–Lewy (CFL) number was set to 0.3, which yielded a time step size of 0.04  ${\rm \mu s}$. The medium was assumed to be lossless and homogeneous with a SOS of 1500 ${\rm m}/{\rm s}$ and a density of 1000 ${\rm kg}/{\rm m}^3$.

Figure \ref{fig:transducer_compare}(a) and \ref{fig:transducer_compare}(b) show spatial maps of the recorded maximum pressure amplitude in a lateral plane, 
corresponding to the numerical model ($p_1$) and reference method ($p_0$). Also,  the spatial relative difference map of the two solutions is shown in Fig. \ref{fig:transducer_compare}(c), where is relative error is less than 0.1\% over all the spatial domain.
Plots of the time-varying pressure signal at a fixed measurement location $\textbf{r}_0$ are compared in Fig. \ref{fig:transducer_compare}(d). 
The pressure signals produced by use of the numerical ($p_1(\textbf{r}_0)$) and reference methods ($p_0(\textbf{r}_0)$) are nearly identical.
These results serve as a validation of the operator $\textbf{H}^{\textbf{u}} \textbf{s}_i$, which is a special case of the forward model when an idealized point receiver is assumed.

To validate the overall forward operator $\textbf{M}_j^{\boldsymbol{\tau}} \textbf{H}^{\textbf{u}} \textbf{s}_i$ for use with lens-focused receiving transducers, a numerical experiment was designed in which the validated operator $\textbf{H}^{\textbf{u}} \textbf{s}_i$ and
the principle of acoustic reciprocity were involved.
Traditionally, the reciprocity principle is stated in terms of a monopole point source and a point receiver; however, the same principle also holds for transducers with an extended aperture and lens curvature\cite{samarasinghe2017acoustic}.
Here, the validation study exploited the fact that the pressure signal measured by the lens-focused transducer in response to a point source is equivalent to the pressure signal produced by the lens-focused transducer and observed at the point source location,
assuming the same excitation source pulse is used.
A simulation was performed that employed an idealized point transducer emitter located at $\textbf{r}_0$. The focused emitter used for the validation of  $\textbf{H}^\textbf{u} \textbf{s}_i$ 
was now employed as a receiver to record the signal produced by the point emitter, which is denoted as $p_2(\textbf{r}_0)$ in Fig. \ref{fig:transducer_compare}(d). 
The close agreement between $p_0(\textbf{r}_0)$ and $p_2(\textbf{r}_0)$ in Fig. \ref{fig:transducer_compare}(d) serves as a validation of the overall transducer model.

\subsection{Characterization via the sensitivity map}
\label{sec:sensi}
To characterize the combined focusing effect of a given emitter-receiver transducer pair, a sensitivity map was computed.
This served to visualize the region of the 3D object that can significantly influence the data recorded by that pair.
The sensitivity map $S(\textbf{r})$ is defined as:
\begin{equation}
\begin{split}
S(\mathbf{r}) \equiv \frac{\int(p_{\mathbf{r}}(t) - p_{0}(t))^2 dt}{\int (p_{0}(t))^2 dt},
\end{split}
\label{eq:Smap}
\end{equation}
where $p_{0}(t)$ is the recorded pressure signal corresponding to the homogeneous medium and $p_{\mathbf{r}}(t)$ is the recorded pressure signal when the point target is inserted at a location $\mathbf{r}$.
As such, the value at each location in the sensitivity map describes how the measured pressure data corresponding to a homogeneous medium would be perturbed by a point target at that location.

\begin{figure}[htb]
\centering
\includegraphics[width=0.48\textwidth]{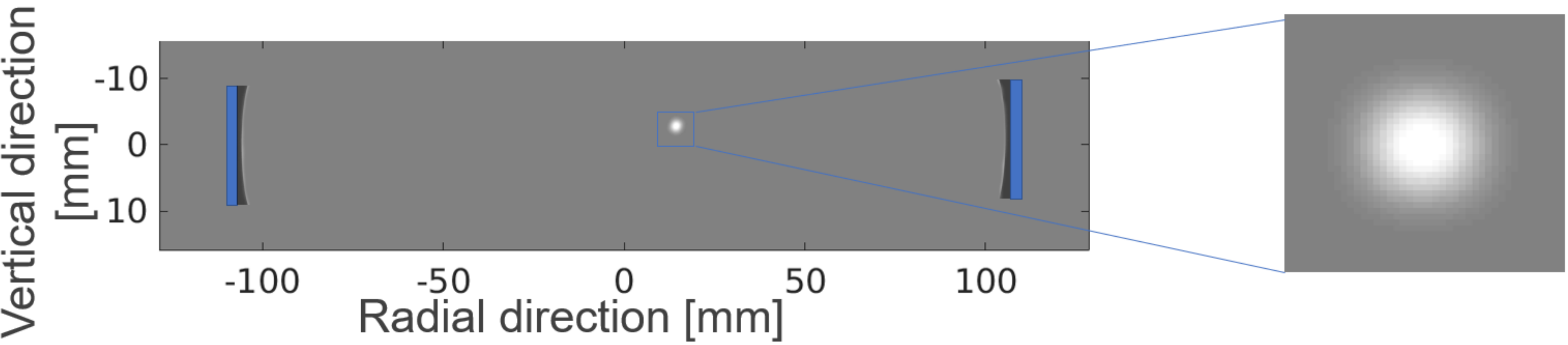}
\vspace{-3.1mm}
\caption{Schematic of the transducer locations and the point target that was employed in the sensitivity map calculation. The transducer emitter-receiver pair is shown as blue rectangles. The point target: a blurred object with a peak SOS value of 1570 m/s.}
\label{fig:sens_geo}
\end{figure}

In this study, the homogeneous medium was defined by the following acoustic parameters: $c_0 = 1500  \text{ m}/{\rm s}$, $\rho_0 = 1000  \text{ kg}/{\rm m}^3$, $\alpha=0$. The inserted point target was modeled as a 3D Gaussian perturbation in the SOS. After the insertion of a point target at location $\textbf{r}$, the SOS distribution was specified as $c_{\textbf{r}}(\textbf{r}') = c_0+c_{1} e^{-\frac{(\textbf{r}'-\textbf{r})^2}{2\sigma^2}}$, where $c_1 = 70 \text{ m}/{\rm s}$ and $\sigma = 0.8 \text{ mm}$.
Figure \ref{fig:sens_geo} shows the relative location of a diametrically opposed emitter-receiver separated by a distance of 220 mm and an illustration of the inserted point target.
The sensitivity map was computed by varying $\mathbf{r}$ within the field of view. 

\begin{figure}[htb]
\centering
\includegraphics[width=0.48\textwidth]{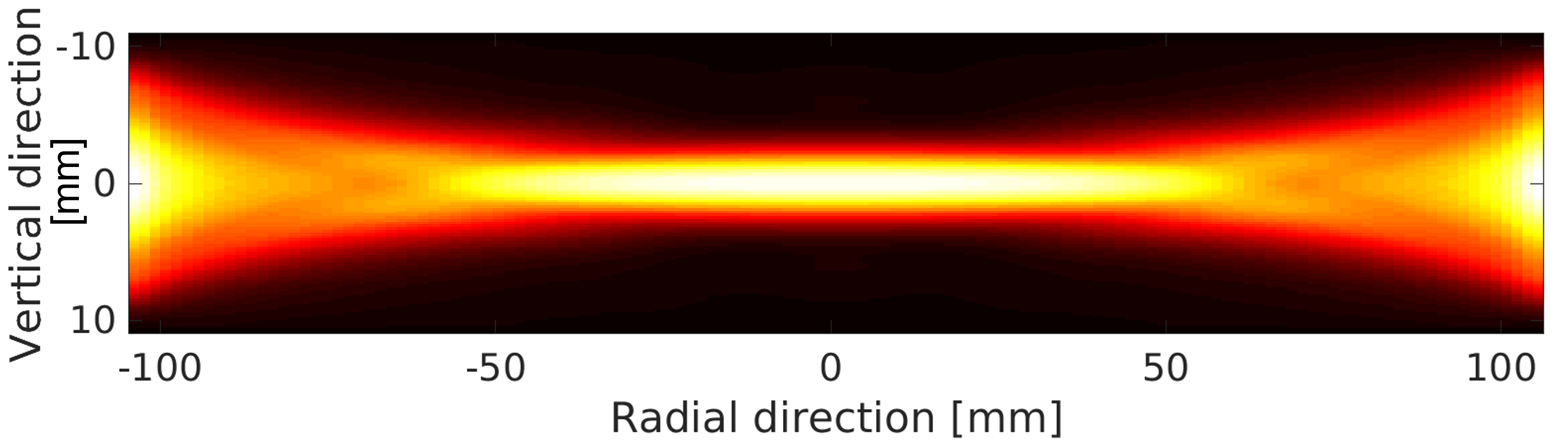}
\vspace{-3.1mm}
\caption{A transverse sensitivity map computed by use of  \eqref{eq:Smap} for a diametrically opposed emitter-receiver pair. This map depicts the elevation-focusing achieved for the emitter-receiver pair.}
\label{fig:sens}
\end{figure}

 Figure \ref{fig:sens} shows the computed sensitivity map. 
 Point targets were uniformly distributed between -10.8 mm and 10.8 mm in the vertical direction with a spatial sampling interval of 0.4 mm, and between -104 mm and 104 mm in the radial direction with a spatial sampling interval of 3.2 mm. The map depicts a thin region (about 6 mm) of high sensitivity to acoustic heterogeneities near the elevation focus.

By enabling visualization of the effective focusing effect achieved by a given emitter-receiver pair, the sensitivity map can provide insights into the vertical resolution of ring-array USCT systems. Additionally, it can guide
the choice of parameters used in the reconstruction method, such as the height of the required computational domain. If measurements acquired at multiple ring-array locations are to be utilized together for image reconstruction, the sensitivity map can also guide the design of the step size by which ring-array should be vertically translated during data-acquisition.


\section{Case study}
\label{sec:cases}
To demonstrate the utility of the proposed forward models for ring-array-based USCT, a virtual imaging study that employed realistic 3D NBPs was conducted. 
This study sought to understand whether 3D image reconstruction of a  slab-shaped volume could yield improved image quality as compared to the conventional 2D SBS approach. The impact of transducer modeling errors on reconstructed 3D images was also investigated.
\label{sec:vis}
\begin{figure}[htb]
    \centering
    \includegraphics[width=0.45\textwidth]{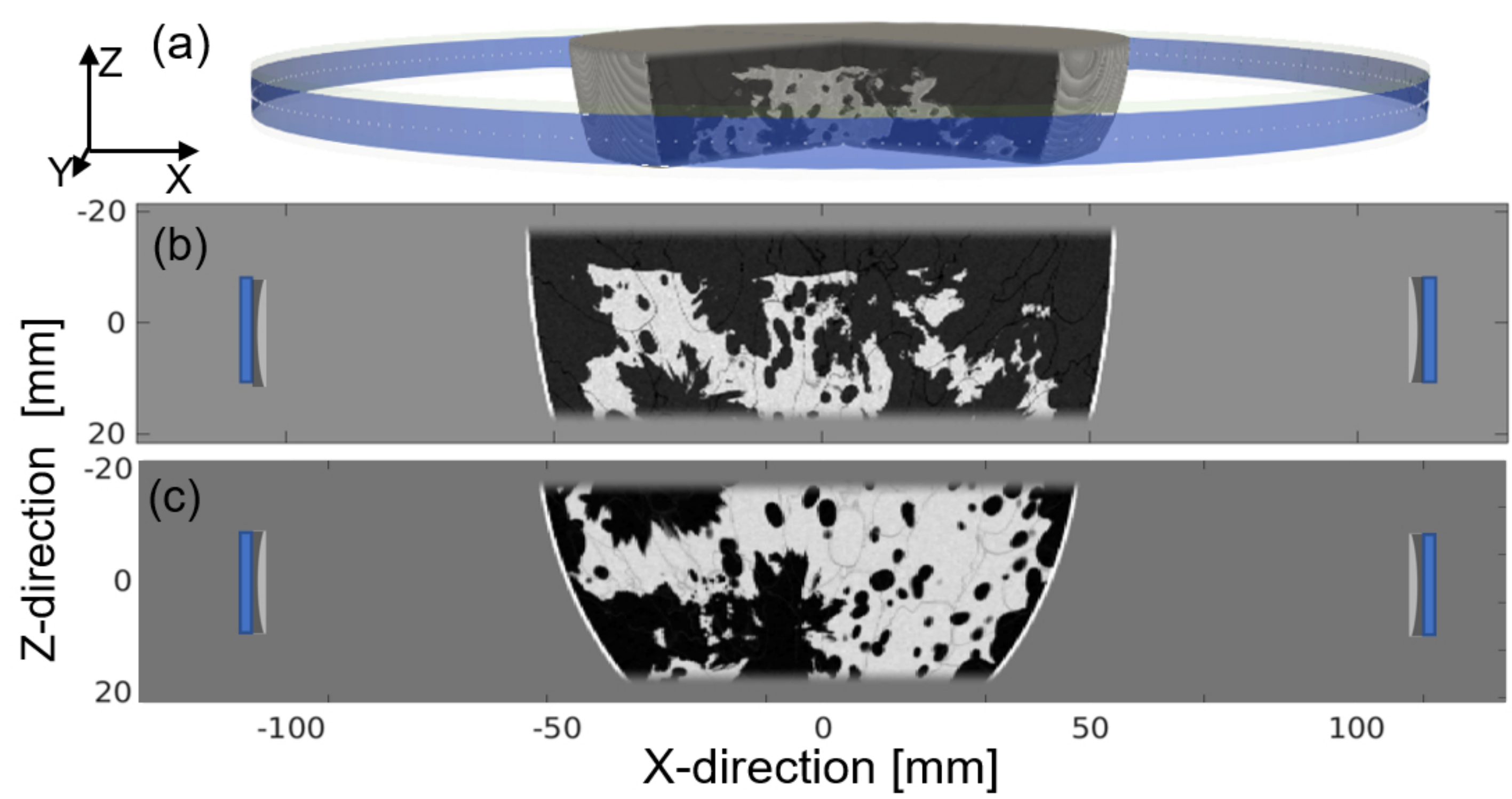}
    \vspace{-4mm}
    \caption{Schematic of the virtual imaging system used in the case study: (a) A 3D rendering of the ring-array USCT system and the object; (b) A transverse plane view of the thin-slab SOS breast phantom of breast type B; and (c) the thin-slab SOS phantom of breast type C.}
    \label{fig:schematic_vit1}
\end{figure}
\subsection{Virtual imaging system}
A virtual ring-array-based USCT system with elevational focusing prescribed by the proposed focused transducer model was assumed.  
The imaging system comprised 128 elevation-focused emitters and 1024 elevation-focused receivers that were evenly arranged in a ring array of a radius of 110 mm \cite{duric2014clinical,9593866}. 
The transducer parameters assumed in Section \ref{sec:validation} were employed. Specifically, a lens with parabolic curvature $a = 0.014 \text{ mm}^{-1}$ (cf. Fig. \ref{fig:source}(b) ) and SOS $c_{\rm lens} = 4500 \text{ m/s}$ was assumed. During data-acquisition, 
measurement data at three different ring-array locations, referred to as ring -1, 0, +1, were acquired by translating the ring-array vertically and stopping at three equispaced locations. The distance between two consecutive scanned imaging planes was 1.8 mm, corresponding to one-tenth of the transducer height. 
In the subsequent reconstruction studies,
the multi-ring data were only utilized to generate a good initial guess for 3D studies.
Only single-ring data from ring 0 were utilized in the final reconstruction. 

Thin slabs extracted from two 3D anatomically realistic NBPs \cite{9537158} corresponding to a BI-RADS breast density category B (scattered areas of fibroglandular density) and C (heterogeneously dense) were virtually imaged. Heterogeneous SOS, AA and density distributions were considered, with realistic values of the tissue properties assigned \cite{9537158}. Figures \ref{fig:schematic_vit1}(b-c) shows crossectional maps of the SOS distribution of the type B and type C NBPs in the transverse plane, respectively. Note that the field of view for the breast type C was chosen near the nipple, resulting in a pronounced vertical curvature.

Further implementation details regarding the wave simulations and measurements data acquisitions are presented in Appendix \ref{app:case_details}.

\subsection{Image reconstruction studies}
\label{sec:case2}

\begin{figure*}[htb]
    \centering \includegraphics[width=0.81\textwidth]{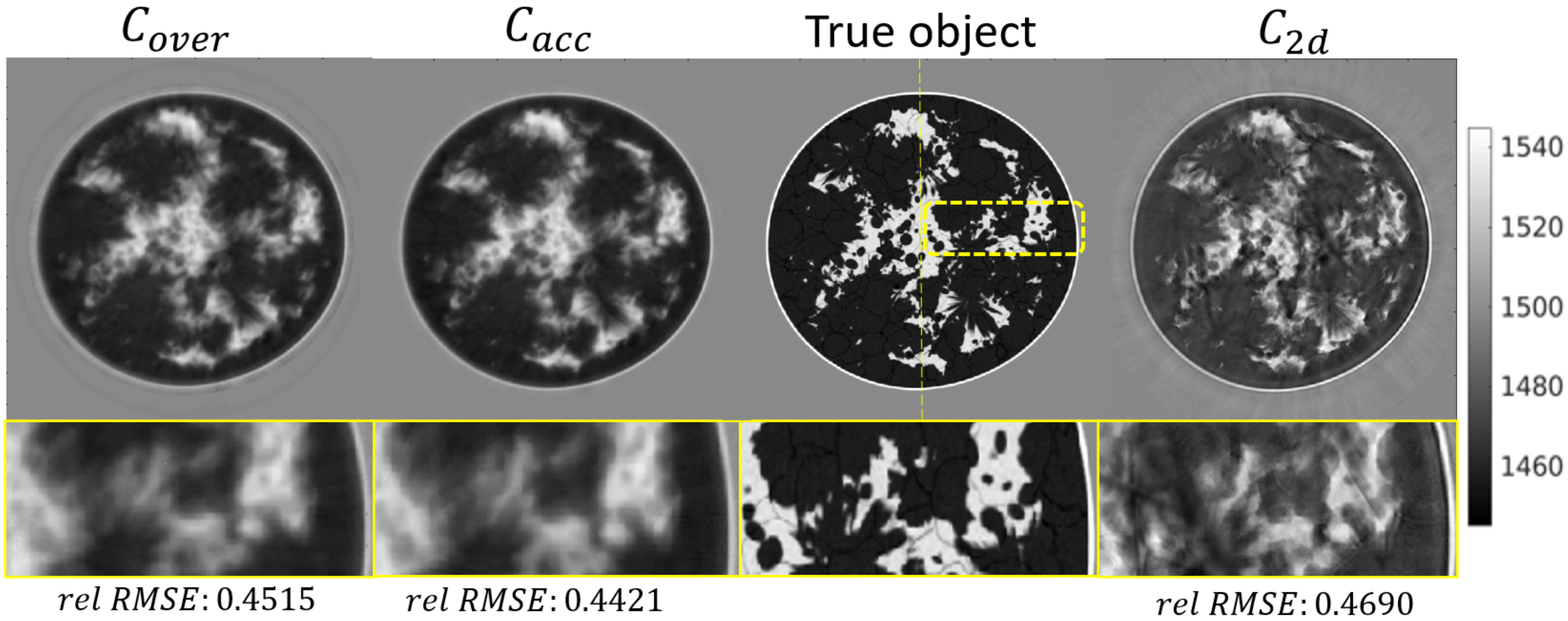}
     \vspace{-3.5mm}
    \caption{Breast type B: Cross-section view of reconstructed SOS distribution by 3D and 2D time-domain FWI at the imaging plane of ring 0. 
    The colorbar is displayed in units of $ \text{m}/\text{s}$. 
    The \emph{rel-RMSE} evaluates the relative error between the slice at imaging plane and the true 2D slice and {\emph{rel-RMSE(c)}$ = \frac{||c - c_{true}||_2}{||c_{0} - c_{true}||_2}$}. 
    The quantities $C_{over}$, $C_{acc}$ denote the object reconstructed by 3D FWI using an overestimated and an accurate transducer model, respectively; $C_{2d}$ denotes the SOS image estimated by 2D FWI. The yellow bounding box in True object indicates the selected region shown of the bottom row images. The vertical yellow dash line indicates the location of the line profiles presented in Fig. \ref{fig:vit2-errmap}.
    }
    \label{fig:vit2-recon}
\end{figure*}

\begin{figure*}[htb]
    \centering \includegraphics[width=0.83\textwidth]{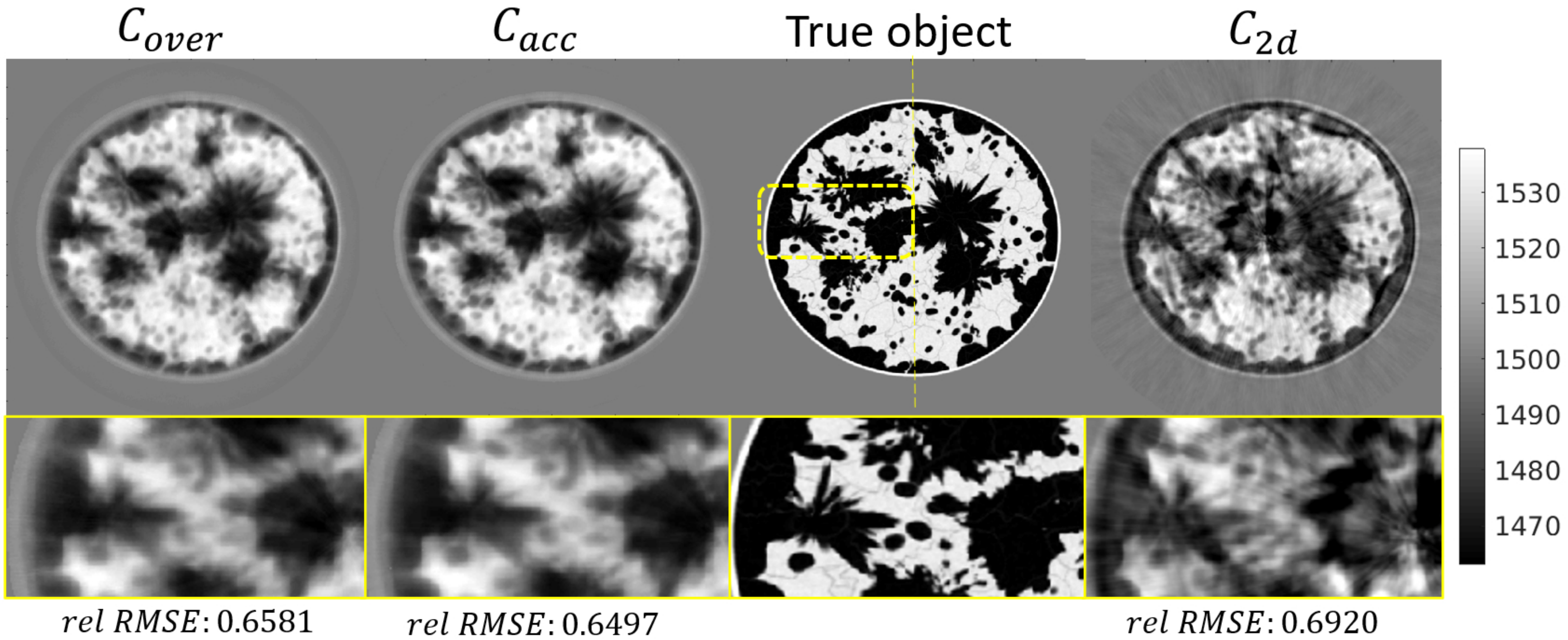}
    \vspace{-3mm}
    \caption{Breast type C: Cross-section view of the reconstructed SOS distribution by 3D and 2D time-domain FWI at the imaging plane of ring 0.   
    The colorbar is displayed in units of $ \text{m}/\text{s}$.}
    \label{fig:vit2-reconh}
\end{figure*}

\begin{figure*}[htb]  
\centering
\includegraphics[width=0.83\textwidth]{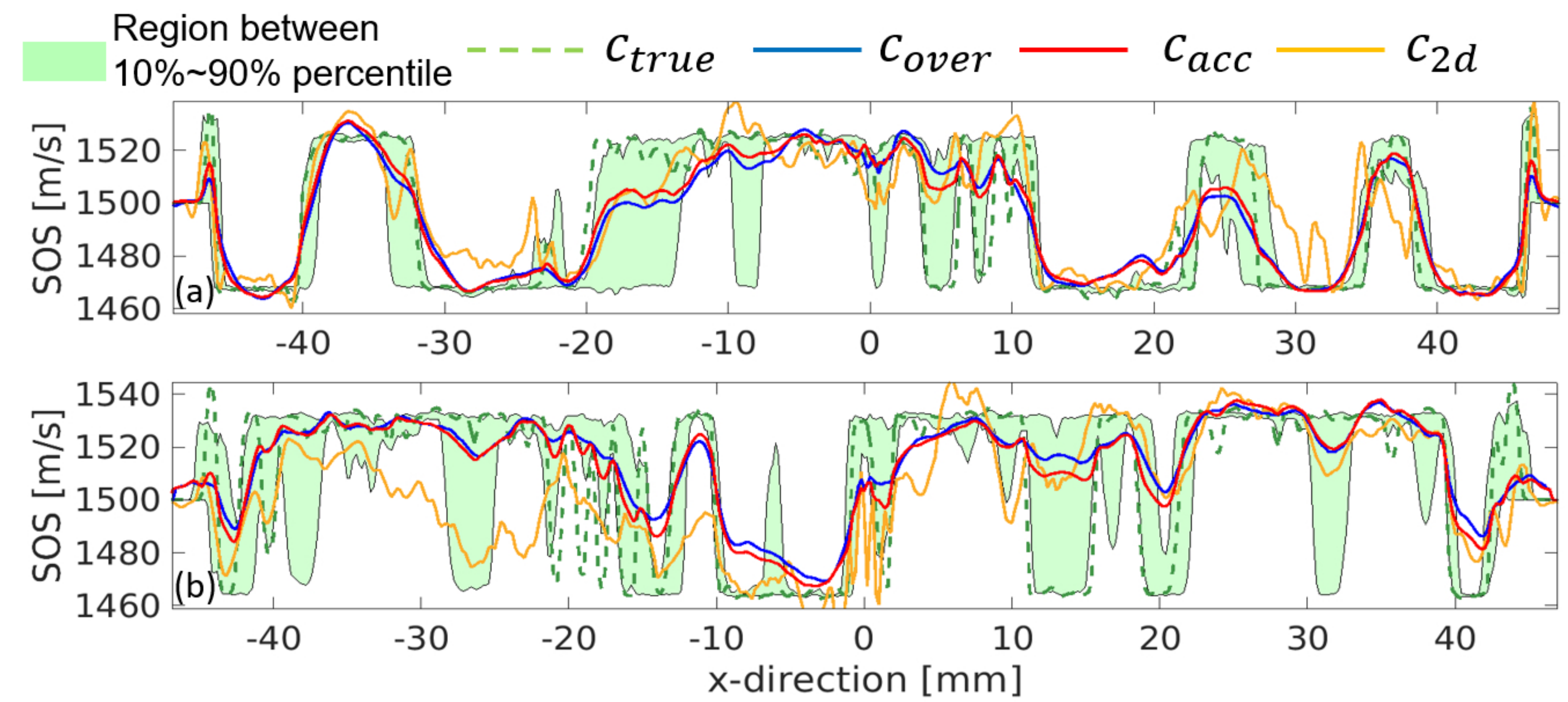}
\vspace{-3.5mm}
    \caption{Line profiles at $y=0$ $mm$ for comparing the estimated SOS using different transducer models. (a) Breast phantom type B; (b) type C. 
    }
    \label{fig:vit2-errmap}
\end{figure*}

The goal of the studies was to investigate the robustness of 3D FWI to transducer modeling errors, as it can be challenging to accurately represent the focusing properties of actual transducers due to their complex design, so that a small modeling mismatch may exist between the actual transducer and the computational model. To this end, slab-shaped SOS volumes were reconstructed using 3D FWI and the proposed forward model, with the exact numerical transducer model used in the forward data generation. An additional 3D reconstruction study was conducted using a transducer model with distorted lens curvature. The transducer model overestimated the curvature of the lens by $10\%$ ($a = 0.0154 \text{ mm}^{-1}$), which can lead to inaccurate time delays and focusing effects.
As a reference, SOS images were reconstructed by 2D FWI using measurement data acquired at ring 0. The transducers were modeled as point-like sources/receivers in the imaging plane. To compare the estimated SOS from the 2D and 3D models, the 2D FWI-reconstructed image was compared to 2D slices of the 3D FWI-reconstructed images at the imaging plane of ring 0. 

In these reconstruction studies, a two-step coarse-to-fine grid method for time-domain FWI was used to address the issue of constructing a good initial SOS model to avoid cycle-skipping and reduce the computational burden. In the first step, a homogeneous SOS medium ($c_0 = 1500 \text{ m}/\text{s}$) was used as the starting model and FWI was performed on a coarser grid using lower frequency measurements, to estimate a good initial guess for finer grid reconstruction.
In the second step, the estimated SOS map from the first step was refined on a finer grid using higher frequency data. Specifically, in all 3D reconstructions, initial guesses were obtained by combining three coarse grid SOS estimates using measurements acquired from rings -1, 0, +1. This approach, which uses multi-ring data, allows for a more accurate SOS representation of the object at different elevation positions, particularly in the high sensitivity region shown in Fig. \ref{fig:sens}. However, only measurements at ring 0 were used in the final finer grid FWI. 
In all reconstructions, a heuristic approach to compensate for the heterogeneity of AA and density media was used. Specifically, a two-region piece-wise constant model of AA/density media was used that assumes that
the breast boundary is known (reflectivity imaging could be possibly used to estimate it) and assigns a constant AA/density value to the background and another to the breast region, which corresponds to a weighted average of the mean values of AA/density in fatty and glandular tissues \cite{9537158}.
The heuristic approach is designed as a fair approach for comparing 2D and 3D methods. The accurate reconstruction of SOS with AA/density mismatches and the impact of such mismatches are not the primary focus of this paper\cite{Multiparametersinv}.
Additionally, measurements were corrupted with Gaussian independent and identically distributed (i.i.d.) noise. 
The implementation details of the two-step reconstruction method, the approach of the initial guess formation from three coarse grid reconstructions, the use of perturbed time delays, and the noise model are provided in Appendix \ref{app:case_details}.

\subsection{Results}
The reconstructed SOS maps at ring 0 are shown in Fig. \ref{fig:vit2-recon} for breast phantom type B, and Fig. \ref{fig:vit2-reconh} for breast phantom type C. 
Line profiles corresponding to the presented 2D SOS maps (at the thin yellow lines) are shown in Fig. \ref{fig:vit2-errmap}.
The green shaded region indicates the vertical variance of true SOS in a high sensitivity region (from -4 mm to 4 mm), where
the upper and bottom boundaries are the 90\% percentile and
10\% percentile of the SOS distribution of each z-column.
The shaded region reveals the large SOS heterogeneity of the employed phantoms along the vertical direction.

\section{Discussion }
\label{sec:dissusion}
The results of the case study show that modeling wave propagation in 3D and accounting for elevation focusing can overcome the limitation of the 2D approach and enable high-resolution estimates of SOS in the imaging plane, even when data from only a single elevation of the ring-array are used.
Particularly, the 2D reconstruction presents significant artifacts and incorrect tissue structures due to the out-of-plane scattering and 2D/3D model mismatch. Notably, these image artifacts are similar to those observed in USCT images obtained from actual experimental data by the 2D method \cite{jw2017}.
The object estimates produced by 3D FWI have a more accurate estimation of tissue structures, SOS values, and lower artifacts level, even when the transducer curvature (and thus its focusing properties) are not known exactly. 
More importantly, as shown in Fig. \ref{fig:vit2-errmap}, the profiles of 3D FWI results mostly reside within the shaded region, whereas the profiles of 2D FWI do not. This show that 3D FWI results can 
provide SOS estimates that, despite the limited spatial resolution in the vertical direction, are consistent with SOS values of the 3D object within the high-sensitivity region of the ring-array. 
The results demonstrated that such a 3D reconstruction approach can improve the accuracy of SOS reconstruction, and it is robust with respect to uncertainty in the transducer model parameters. 

However, mismatches clearly exist between the true object and the object estimated by 3D FWI. Also, the 3D FWI results appear smoother compared to the 2D FWI. These are expected since only single-ring measurements were used in the finer grid reconstruction, which results in an underdetermined 3D reconstruction problem. This
is the cause of the limited spatial resolution of the images. On the other hand, the single-ring measurement data are sufficient for 2D image reconstruction; however substantial modeling errors arise by applying a 2D
imaging model to reconstruct data generated via 3D wave propagation physics and lens-focusing. As such, the images reconstructed by use of the 2D method appear sharp but contain conspicuous artifacts due to the
modeling error involved.
This motivates the development of new algorithms incorporating USCT data corresponding to multi-position of the ring-array to further improve the spatial resolution of 3D reconstruction\cite{li3dfwi}. 

The proposed model possesses a few limitations. First, the in-plane focusing (directivity) of the transducers is not modeled due to the high aspect ratio of the transducers employed in ring-array USCT. Additional discussions about modeling the directivity in ring-array USCT, especially when the in-plane aperture is not significantly larger than the pixel size, can be found in the literature \cite{wise2019representing,yuan2023full}. Second, the transducer model is Cartesian grid-based and requires the use of nearest-neighbor interpolation when the transducer location does not coincide with the voxel size. 
Non-Cartesian grids can also be explored as potential alternatives.  One such approach is the isogeometric finite element method\cite{cottrell2009isogeometric}, which allows for an unstructured grid that can conform to the curved shape of the transducer\cite{komatitsch2018specfem,schoeder2019exwave}. Nevertheless, it has challenges including computational cost and implementation difficulties, particularly when it comes to efficient hardware parallelizations.
To reduce the modeling errors due to the irregular spacing in the non-Cartesian grid inside the transducer, one potential solution is applying varying weights scaling factors to different grid points.
Third, the effects of the electrical impulse response (EIR) are not considered. However,  the EIR can be readily incorporated as described elsewhere\cite{5560859}. 
Fourth, in the employed lens-focused transducer model, the attenuation of the lens and the acoustic impedance mismatches at the lens surface are not taken into account. Related discussions are presented in Appendix \ref{app:impedence}.

It is also worth noting that, in practical applications, lens parameters should be carefully designed for the desired focusing performance. An important consideration is the trade-off between the desired focusing performance and the loss induced by the lens (due to AA and acoustic impedence changes as described in Appendix \ref{app:impedence}) as this could reduce the signal-to-noise ratio. In addition,
The SOS value and curvature profile of the lens in simulation studies may not be optimal. Nevertheless, these  are parameters of the proposed transducer model that can be adjusted as needed.

However, accurate transducer modeling for real experimental data remains a challenging task. 
To reduce transducer modeling errors for practical use, one approach involves calibration, which treats the excitation source pulse, transducer aperture, and time delays as a surrogate model. Further discussions on this topic can be found in the literature \cite{9512083,roy2011robust}. 

Finally, in this study, only in-plane data acquisitions were employed to emulate the existing ring-array USCT system. However, the proposed model can be extended for more flexible cases.  For instance, it can be adapted to model the case where receivers are positioned at different heights or where a vertically rotated ring-array configuration is utilized. Thus, the proposed model can be also used to explore new system geometries that could potentially offer improved vertical solutions.



\section{Summary}
\label{sec:conclusion}
In this study, a 3D forward model on a regular
Cartesian grid accounting for the elevation focusing effects in ring-array USCT systems was presented to enable accurate ultrasound simulation of the USCT data acquisition process. The focusing effects are modeled by decomposing each transducer into a discrete collection of source/receiver patches and applying a spatially varying time delay to the ultrasound pulse (emitter mode) and recorded signal (receiver mode).
The proposed numerical transducer model was quantitatively validated by the use of a semi-analytical approximation of the Rayleigh-Sommerfeld integral solution and of the acoustic reciprocity principle.
The case study demonstrated that, by accounting for transducers' focusing properties and 3D wave propagation effects, the proposed 3D imaging model can lead to more accurate estimates of SOS in the imaging plane and fewer artifacts compared to estimates obtained using a 2D wave propagation model, such as that used in the SBS reconstruction approach. Furthermore, the case study demonstrated the robustness of 3D FWI-based image reconstruction using the proposed imaging model with respect to uncertainty in the focusing properties of the transducer, which may be difficult to accurately characterize experimentally.

This work is a key step towards enabling improved volumetric image reconstruction in elevation-focused ring-array USCT.
It also provides a foundation for conducting 3D image reconstruction studies that employ data acquired at multiple ring-array positions.

\appendices
\section{Publicly released datasets}
The NBPs and computer-simulated pressure traces used in the case study have been released under Public Domain CC-0 dedication and are available for download from \cite{li2021NBPs3D}.


\section{Modeling of acoustic impedance discontinuity and attenuation by apodization weight function}
\label{app:impedence}
In a practical lens-focused transducer, discontinuities in acoustic impedance often occur at the interface between the lens and the propagation medium. Additionally, AA is often present in the lens materials. 
In this section, the modeling of a lens-focused transducer taking into account the acoustic impedance and AA is presented by introducing an apodization weight function. Additional simulation studies are conducted to demonstrate the use of the apodization weights together with the proposed transducer model in a scenario of practical relevance.
\begin{figure}[htb]
    \centering
       \vspace{-3mm}
    \includegraphics[width=0.39\textwidth]{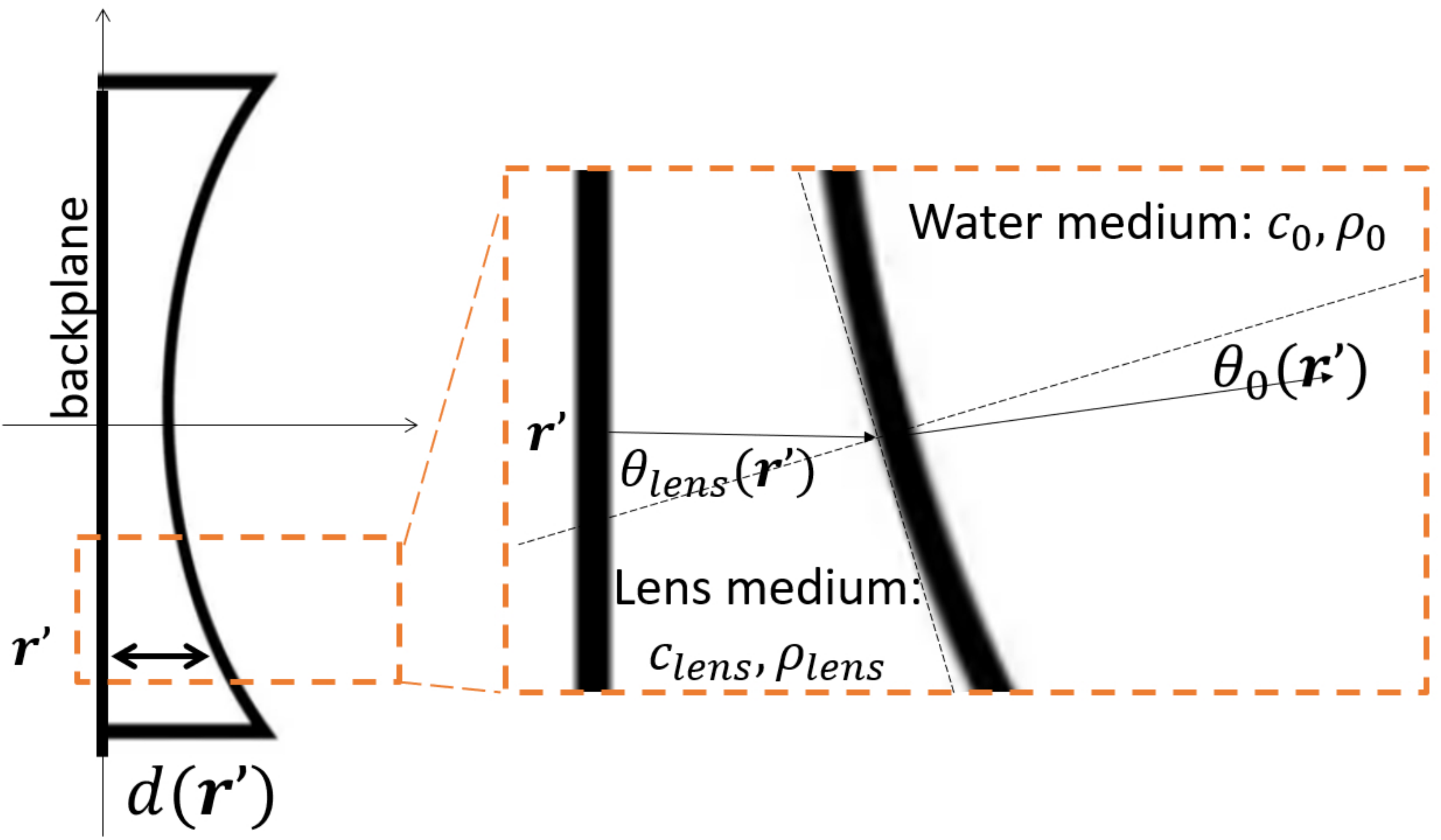}
   \vspace{-4mm}
    \caption{Illustraction of lens thickness $d(\textbf{r}')$ and incident angle $\theta_{lens}$ and refraction angle $\theta_{0}$.}
    \label{fig:theta}
\end{figure}

For the transducer in emitter mode, the apodization weight function is applied to the normal velocity, such that \eqref{eq:lens_pvel} is updated as \cite{911727, marechal2007lens}:
\begin{equation}
    v_{\bot } \left(\textbf{r}', t \right) = \hat{v}\left(t+\tau(\textbf{r}') \right) \, w_e(\textbf{r}')  \quad \forall \textbf{r}' \in \mathbf{\Omega},
    \label{eq:lens_pvel2} 
\end{equation}
where $w_e(\textbf{r}') $ denotes the apodization weight function at the emitter. It accounts for the attenuation loss due to the AA within the lens and reflection loss due to impedance mismatch as waves travel from the lens to a homogeneous medium. Specifically $w_e(\textbf{r}')=a(\textbf{r}')z_e(\textbf{r}')$, where $a(\textbf{r}')$ is the attenuation factor, which can be modeled as a function of lens thickness $d(\textbf{r}')$ and the AA value of the lens using a ray-based approximation, and $z_e(\textbf{r}')$ is the transmission coefficient accounting for the reflection loss. Assuming the lens is a fluid medium, the transmission coefficient $z_e(\textbf{r}')$ for the transition from the lens to a water medium can be derived based on the laws of reflection and refraction \cite{marechal2007lens,dukhin2010fundamentals}:

\begin{equation}
\label{eqn:transmission}
\begin{aligned}
z_e(\textbf{r}') &= \frac{2c_{0}\rho_0 \cos\theta_{lens}}{c_{0}\rho_0 \cos\theta_{lens}+c_{lens}\rho_{lens} \cos\theta_0};\\[2pt]
&\text{ with } \quad \frac{\sin \theta_{lens}}{c_{lens}} = \frac{\sin \theta_0}{c_{0}},
\end{aligned}
\end{equation}
where $\rho_{lens}$ and $\rho_{0}$ represent the density values in the lens and homogeneous medium, respectively. The variables $\theta_{lens}$ and $\theta_0$ are functions of $\textbf{r}'$, denoting the incident angle and refraction angle, as shown in Fig \ref{fig:theta}. 

Similarly, for the transducer in receiver mode, \eqref{eqn:c_receiver} is updated as:
\begin{equation}
\begin{aligned}
 g_{\mathbf{\Omega}}(t) &= \iint_\mathbf{\Omega} w_r(\textbf{r}') p\left(\textbf{r},t+\tau(\textbf{r}) \right) d \textbf{r},
\label{eqn:c_receiver2} 
\end{aligned}
\end{equation}
where $w_r(\textbf{r}')$ is the apodization weight function at the receiver. It is modeled as $w_r(\textbf{r}') = a(\textbf{r}') z_r(\textbf{r}')$,
where $z_r(\textbf{r}') = 2- z_e(\textbf{r}')$ accounts for the loss when the wave travels from the propagation medium to the lens.

\begin{figure}[htb]
\centering
\vspace{-3mm}
\includegraphics[width=0.47\textwidth]{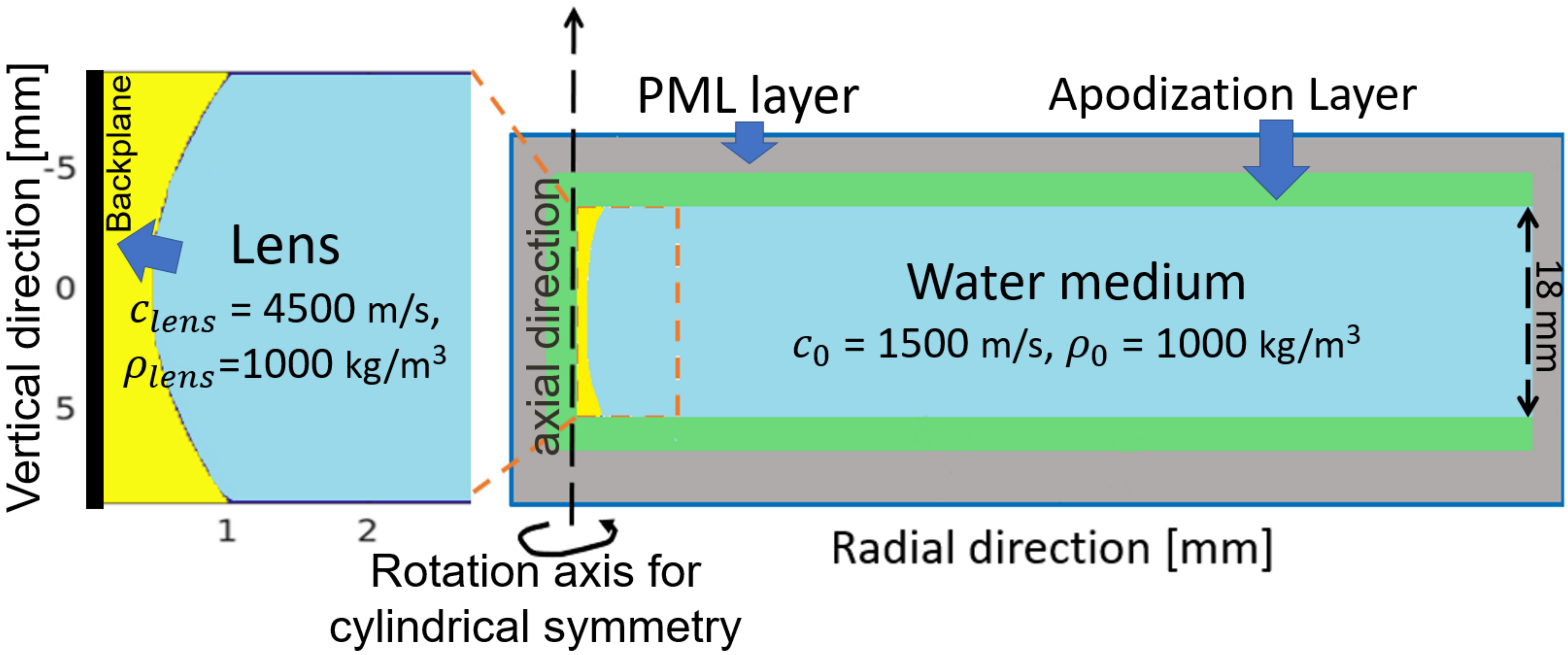}
\vspace{-3.1mm}
\caption{Illustration of a 2D axisymmetric wave simulation medium. Here, the impedance of the lens is 3 times greater than the impedance of the water medium. }
\label{fig:ASsimulation}
\end{figure}
To validate the proposed apodization weight function, and in particular, the acoustic impedance correction $z_e(\textbf{r}')$, an additional simulation study was conducted. 
A non-attenuating lens was assumed in this study. A high-resolution 2D axisymmetric simulation\cite{2das} is utilized as a reference. The simulation medium is shown in Fig. \ref{fig:ASsimulation}, where the lens is explicitly constructed within the propagation medium by assigning different SOS values to the lens region. In this case, an acoustic impedance discontinuity is present between the lens and propagation medium, where $c_{lens}\rho_{lens} = 3 c_{0}\rho_{0}$.
A non-delayed excitation source pulse is assigned to each pixel on the backplane. The phase changes are achieved by the higher SOS in the lens region. The generated wavefield data is then compared with the semi-analytical integral solution incorporating the apodization weight function (with transmission coefficient only), based on  \eqref{eqn:solution1}\eqref{eqn:timedelay}\eqref{eq:lens_pvel2}\eqref{eqn:transmission}.

The proposed 3D transducer model exhibits cylindrical symmetry and monopolar characteristics (i.e., no directivity) in the imaging plane. As a result, simulation of such transducer model can be efficiently represented by a 2D axisymmetric simulation, which can largely reduce computational costs and allows for simulations on a finer grid. The computational simulation domain was a high-resolution 2D Cartesian grid with 500 × 5600 pixels. Each pixel had a size of 0.04 mm. Accordingly, the transducer was divided into a 1D array of 450 pixels. To ensure numerical stability, the Courant–Friedrichs–Lewy (CFL) number was set to 0.2, resulting in a time step size of 2 ns. An apodization layer (cf. Appendix \ref{app:generation}) was introduced between the lens and the perfectly matched layer (PML) \cite{berenger1994perfectly} to mitigate strong reflections at the back, top, and bottom edges of the lens due to acoustic impedance mismatches.

\begin{figure}[htb]
\centering
\vspace{-3mm}
\includegraphics[width=0.4\textwidth]{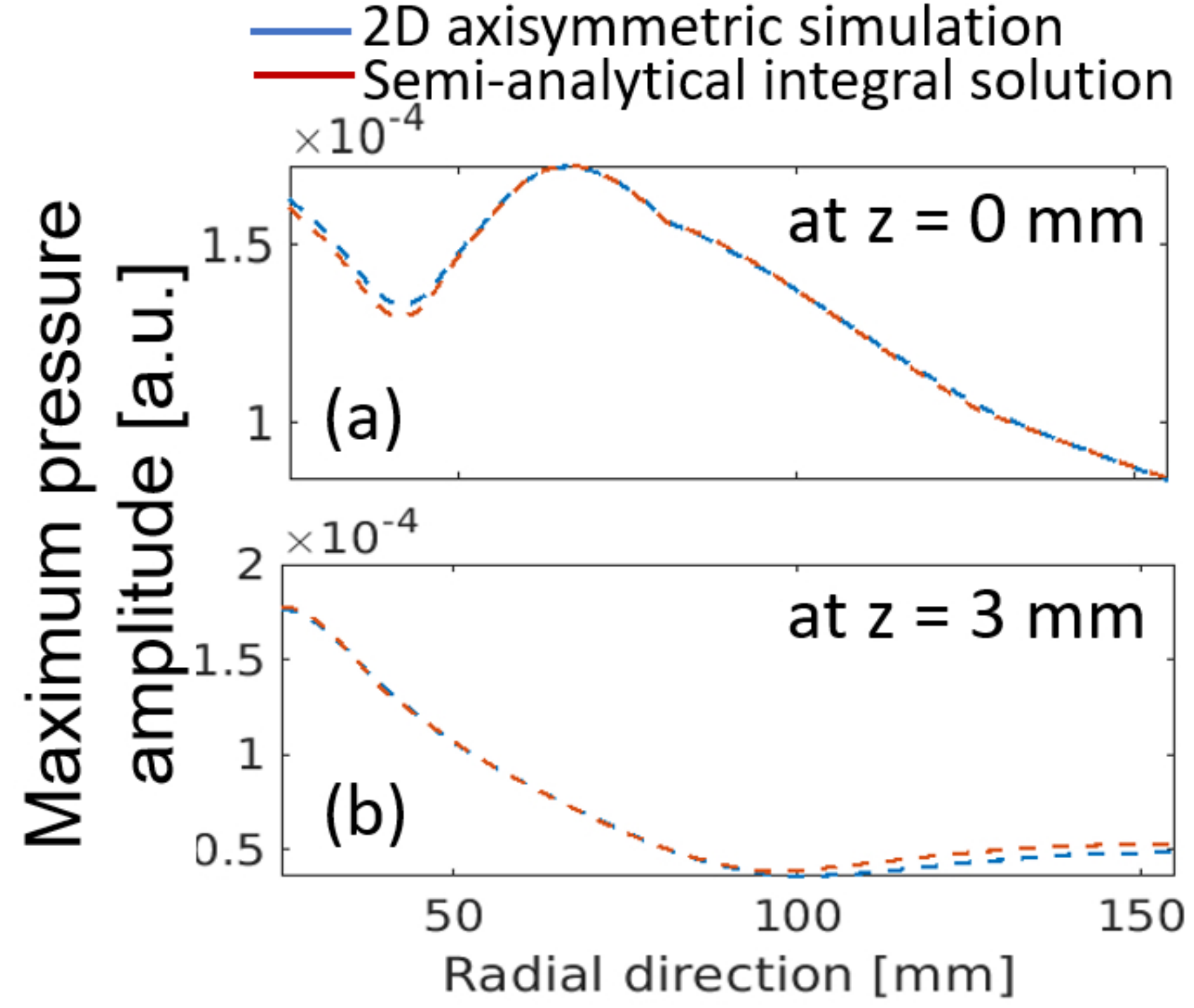}
\vspace{-3.5mm}
\caption{Maximum pressure amplitude profiles for 2d axisymmetric simulation and semi-analytical integral solution with apodization weight function. Lines along the radial direction were plotted (at z = 0 mm and z = 3 mm). The excitation pulse used in the computation is shown in  Fig \ref{fig:source}(a). Pressure amplitude profiles are reported in the same unit as the those used in Fig. \ref{fig:transducer_compare}(d).  }
\label{fig:appb_profile}
\end{figure}

Figure \ref{fig:appb_profile} shows the comparison of the generated pressure data by two methods.  A reasonably close agreement between the two solutions was observed, demonstrating the effectiveness of the apodization weight function in mitigating the impedance mismatch. 
However, it is important to clarify that
we did not anticipate a perfect match, as the 2D simulation is a discretized model with small stair-casing errors when representing the continuous curvature of the lens on a discrete grid.
In addition, due to the complicated design of transducers, 
accurate modeling of lens-focused transducers in real applications remains a challenging task. The proposed lens model is still a simplified computational model and does not account for the presence of one or more matching layers typically used in piezo-electric transducers to reduce losses induced by acoustic impedance mismatch. 
To reduce the modeling errors, the apodization weight function can be treated as an unknown surrogate variable and calibrated using experimental data \cite{9512083,roy2011robust}.

\section{Implementation details of the case study}
\label{app:case_details}

\subsection{Generation of the measurement data}
\label{app:generation}
\begin{figure}[htb]
    \centering
    \includegraphics[width=0.49\textwidth]{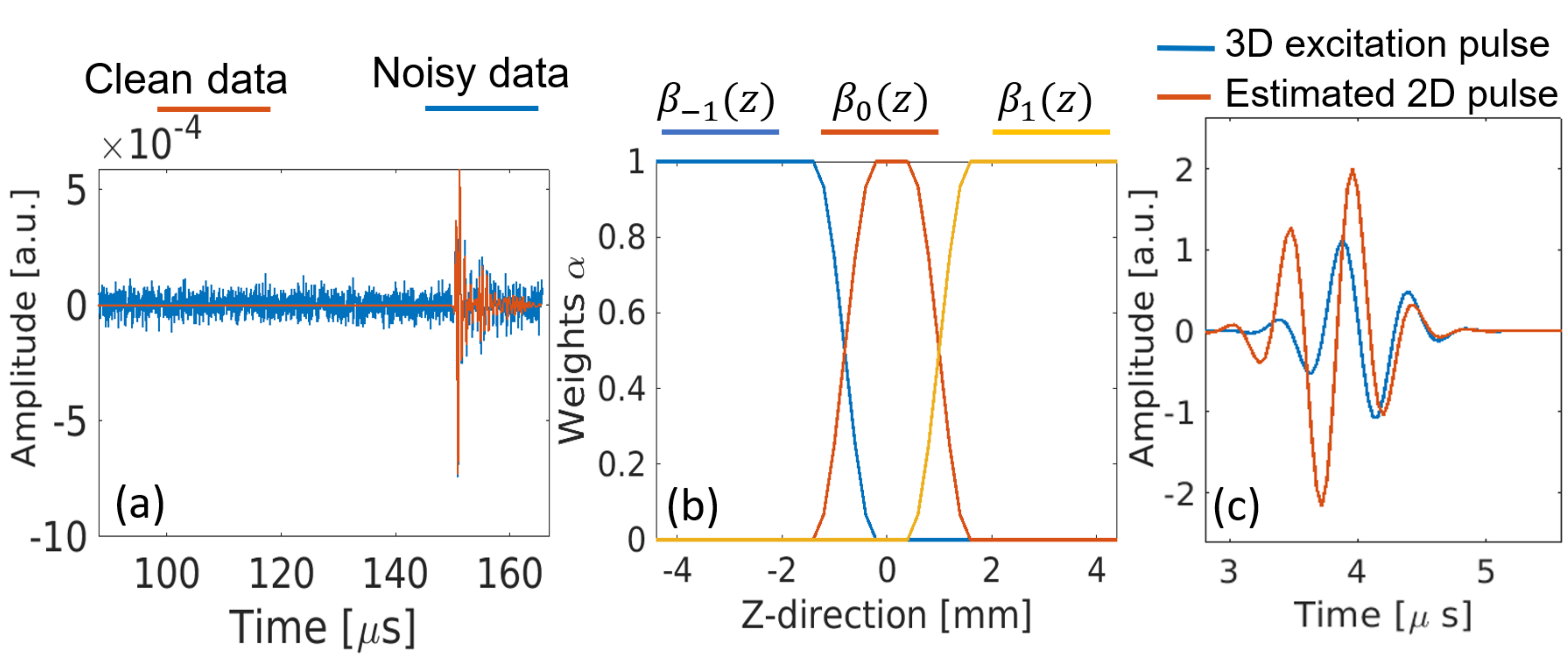}
    \vspace{-4.5mm}
    \caption{(a): Profiles of the clean and noisy data emitted by the 1-th emitter and received by the 512-th receiver, for \textit{SNR}$_{dB} = 22$ $dB$. (b): The profile of weights $\beta_i(z)$ for combining coarser grid reconstructions from three different ring measurements. (c): The 3D excitation pulse and the estimated effective 2D excitation pulse. }
    \label{fig:noise-signal}
\end{figure}

The forward simulation of the USCT measurement data was conducted using k-Wave\cite{treeby2010k}, an open-source toolbox for time-domain ultrasound
simulations based on the k-space pseudospectral method.  To avoid strong wave reflections and scattering that are due to the use of a truncated thin slab phantom and that are not present in a clinical setting, the computational domain is embedded in an extended domain that includes an apodization layer and PML.
In the apodization layer, heterogeneous SOS and density maps of the to-be-imaged object are smoothly apodized to match those of the homogeneous coupling medium (water). This apodization layer is introduced because PML can effectively suppress wave reflection at the boundaries of the computational domain \cite{berenger1994perfectly,treeby2010k} only when acoustic heterogeneities are bounded away from the boundary of the domain. In the case study, the computational domain has a height of 27.2 mm and it is embedded in an extended domain with height 43.2 mm by the addition of the apodization layers (4 mm each) and PMLs (4 mm each).
The discretization parameters of the forward wave simulation are reported in Table \ref{tab:systems_para2}.
For all reconstruction studies, measurement data were corrupted with additive Gaussian noise $\boldsymbol{\epsilon}$ with components that are assumed to be independent and identically distributed (i.i.d.) with a zero mean and a standard deviation of 3.25e-4. This results in a signal-to-noise ratio (SNR) of 22 dB, which is the logarithmic ratio between the power of the noiseless signal and the power of the noise.
An example of time trace of noisy pressure data is shown in Fig. \ref{fig:noise-signal}(a).

\begin{table}[htp!] 
\centering
\caption{Discretization parameters of the virtual imaging system}
\vspace{-2.5mm}
\begin{tabular}{| l | l |}
\hline
Number of emitter/receivers & 128/1024 \\
\hline
3D Computational grid & [1280, 1280, 216]  \\  
 \hline
 Radius of ring-array system & 110 mm \\
 \hline
 Number of ring measurements & 3  \\
\hline
Scanning interval & 1.8 mm  \\
\hline
 Voxel size & 0.2 mm \\
 \hline
  Time step size &  1/25 $\mu$s; CFL number=0.3  \\
  \hline
  Simulation time horizon &  168 $\mu$s\\
  \hline
  Transducer height & 18mm (90 voxels) \\
  \hline
  Central frequency of source pulse & 2 MHz \\
  \hline
  \end{tabular}
  \label{tab:systems_para2}
\end{table}

\subsection{Multi-grid approach for FWI reconstruction in 3D}
Details regarding the implementation of the coarse-to-fine grid FWI image reconstruction approach are described below. This multi-grid approach  employs a lower-resolution lower-frequency FWI reconstruction to provide an initial guess for the higher-resolution higher-frequency reconstruction. Doing so, not only circumvent the cycle-skipping phenomenon but also leads to significant computational savings since most FWI iterations are performed on the coarse grid. In fact, when a coarse grid with voxel size three times larger than the fine grid is used, each coarse grid FWI iteration is approximately 81 times less computationally expensive than a fine grid iteration.
Notably, this multi-grid approach also allows for the use of a larger field of view for the coarse grid reconstruction. As described in \eqref{eqn:sos_gradient}, the computation of the gradient requires access to the forward and adjoint pressure variables at every voxel in the field of view and at every time step. Therefore, computing the gradient for the entire breast region is unfeasible on the fine grid. As a result, while it is feasible to store all pressure variables within the computational domain for the coarse grid reconstruction, in the fine grid reconstruction, only pressure variables within a thin field of view (less than 1 cm of vertical thickness) close to the imaging plane can be stored for gradient computation.

\subsubsection{Coarse grid FWI reconstruction}
In the first step of the reconstruction, the computational grid used for wave simulations is [480,480,108], with a voxel size of 0.6 mm. The transducer is modeled as 30 voxels in height, and the temporal step size is 0.12 $\mu$s.
A low-pass Hann window filter with a cutoff frequency of 0.45 MHz was applied to the excitation pulse and measured pressure traces to prevent cycle-skipping. The height of the field of view included the entire computational domain and apodization layers (35.4 mm).

\subsubsection{Construction of the initial guess for fine grid reconstruction} Three image reconstructions using measurements acquired from three distinct vertical positions of the ring array where performed on the coarse grid to define the initial guess for the fine grid reconstruction. Specifically, for $i \in \{ -1, 0, 1\}$ the vertical position of the ring array is denoted by $z_i = i \Delta z$, where $\Delta z = 1.8 \text{ mm}$ denotes the distance between consecutive vertical positions of the ring array. The coarse grid SOS estimate using data from the i-th ring location is denoted by $c_{i}(x,y,z)$. The initial guess for the fine grid reconstruction was then defined as a weighted sum $c(x,y,z)= \sum_{i \in \{-1, 0, +1\}} c_{i}(x,y,z)\beta_i(z)$ of the coarse grid SOS estimates $c_{i}(x,y,z)$. The weight functions $\beta_i(z)$, those profiles are shown in Fig. \ref{fig:noise-signal}(b), form a partition of unity ($\beta_i(z) \geq 0$ and $\sum_{i \in \{-1,0,1\} } \beta_i(z) = 1$) and are such that $\beta_i(z_j) = \delta_{ij}$, where $z_j$ denotes the elevation of the $j$-th ring ($j\in\{-1,0,1\}$) and $\delta_{ij}$ is a Kronecker delta function.

\subsubsection{Fine grid FWI reconstruction}
In the second step, the computational grid size is [1280,1280,180]. The voxel size (0.2 mm) and temporal step size (0.04 $\mu$s) is the same as the one in the forward simulation. 
A low-pass filter with a cutoff frequency of 1.2 MHz is applied to the excitation pulse and measurement data.  The height of the field of view is 8 mm that corresponds to a thin slab of high sensitivity to acoustic heterogeneity centered at the imaging plane (cf. Fig. \ref{fig:sens}).
\subsubsection{Stopping criteria}
The stopping criteria for this multi-grid strategy are defined as follows: The iteration with minimum \textit{root-mean-square errors} (\textit{RMSE}) with the true object was selected in both the coarse and fine grid reconstructions.

It is important to note that stopping rules based on the minimum RMSE are not suitable for
practical applications since the true object is unknown. However, the primary focus of this work is forward (transducer) modeling and not the development of a practical image reconstruction method.  This simple stopping criteria was adopted for the purpose of eliminating the choice of more sophisticated stopping rules
for the 2D and 3D studies as a potential confounding factor. By minimizing the RMSE, the comparison of 2D
and 3D methods could be performed in a fair way in our computer-simulation studies where the true objects were known.
The exploration of alternative stopping criteria or various optimization methods that may improve image quality is left for future research.

\subsection{Details on the reference 2D FWI reconstruction}
The reference SOS image was reconstructed by 2D FWI using measurement data acquired at ring 0. The transducers were modeled as point-like sources/receivers in the imaging plane. 
To mitigate the discrepancy between wave propagation in 2D and 3D, an effective excitation pulse was estimated by minimizing the Euclidean norm of the difference between pressure traces measured by a receiver diametrically opposed to the source when 2D and 3D wave propagation through a homogeneous medium (water). The 3D excitation pulse and the estimated effective 2D excitation pulse are shown in Fig. \ref{fig:noise-signal}(c).

The reference SOS image was obtained by adapting the above FWI multi-grid approach to the 2D case.
First, an initial guess is estimated by FWI on a 2D coarser grid with 480$\times$480 pixels, 0.6 mm pixel size, 0.12 $\mu$s temporal step size. A low-pass Hann window filter with a cutoff frequency of 0.45 MHz was applied to the excitation pulse and measured pressure traces.
The final reconstruction was then performed on a finer grid with 1280$\times$1280 pixels, 0.2 mm pixel size, 0.04 $\mu$s temporal step size. A low-pass filter with a cutoff frequency of 1.2 MHz is applied to the excitation pulse and measured pressure traces. The same stopping criteria employed in the 3D method is used in 2D. 

\subsection{Computational time}
The majority of the reconstruction time is dedicated to the fine-grid reconstruction process. In the case of 3D reconstruction, it takes approximately 2.3 days using a single NVIDIA A100 (40GB memory) GPU on the Delta supercomputer at the University of Illinois Urbana-Champaign. Each iteration of the 3D fine grid FWI requires around 34 minutes, while each iteration of the 3D coarse grid FWI takes about 50 seconds.

For the 2D reconstruction, the process takes approximately 4.5 hours using a single NVIDIA TITAN X Pascal (12 GB memory) GPU. Each iteration of the 2D fine grid FWI takes 120 seconds. On the coarser grid, each iteration takes 16 seconds.

\section*{Acknowledgment}
 Computational resources for this work were granted to the authors by the Delta research computing project, which is supported by the National Science Foundation (award OCI 2005572), and the State of Illinois. Delta computing project is a joint effort of the University of Illinois at Urbana-Champaign and its National Center for Supercomputing Applications.





\ifCLASSOPTIONcaptionsoff
  \newpage
\fi

\bibliography{usct} 

\begin{thebibliography}{10}
\providecommand{\url}[1]{#1}
\csname url@samestyle\endcsname
\providecommand{\newblock}{\relax}
\providecommand{\bibinfo}[2]{#2}
\providecommand{\BIBentrySTDinterwordspacing}{\spaceskip=0pt\relax}
\providecommand{\BIBentryALTinterwordstretchfactor}{4}
\providecommand{\BIBentryALTinterwordspacing}{\spaceskip=\fontdimen2\font plus
\BIBentryALTinterwordstretchfactor\fontdimen3\font minus
  \fontdimen4\font\relax}
\providecommand{\BIBforeignlanguage}[2]{{%
\expandafter\ifx\csname l@#1\endcsname\relax
\typeout{** WARNING: IEEEtran.bst: No hyphenation pattern has been}%
\typeout{** loaded for the language `#1'. Using the pattern for}%
\typeout{** the default language instead.}%
\else
\language=\csname l@#1\endcsname
\fi
#2}}
\providecommand{\BIBdecl}{\relax}
\BIBdecl

\bibitem{andre2013clinical}
M.~Andr{\'e}, J.~Wiskin, and D.~Borup, ``Clinical results with ultrasound
  computed tomography of the breast,'' in \emph{Quantitative ultrasound in soft
  tissues}.\hskip 1em plus 0.5em minus 0.4em\relax Springer, 2013, pp.
  395--432.

\bibitem{roy2013breast}
O.~Roy, S.~Schmidt, C.~Li, V.~Allada, E.~West, D.~Kunz, and N.~Duric, ``Breast
  imaging using ultrasound tomography: From clinical requirements to system
  design,'' in \emph{2013 IEEE International Ultrasonics Symposium
  (IUS)}.\hskip 1em plus 0.5em minus 0.4em\relax IEEE, 2013, pp. 1174--1177.

\bibitem{guasch2020full}
L.~Guasch, O.~Calder{\'o}n~Agudo, M.-X. Tang, P.~Nachev, and M.~Warner,
  ``Full-waveform inversion imaging of the human brain,'' \emph{NPJ digital
  medicine}, vol.~3, no.~1, pp. 1--12, 2020.

\bibitem{wiskin2020full}
J.~Wiskin, B.~Malik, D.~Borup, N.~Pirshafiey, and J.~Klock, ``Full wave 3{D}
  inverse scattering transmission ultrasound tomography in the presence of high
  contrast,'' \emph{Scientific Reports}, vol.~10, no.~1, pp. 1--14, 2020.

\bibitem{wiskin2022imaging}
J.~W. Wiskin, J.~Enders, I.~Turkbey, M.~Rothberg, M.~Merino, S.~Xu, E.~Boctor,
  Y.~Wu, A.~Negussie, J.~Klock \emph{et~al.}, ``Imaging of prostate cancer with
  3{D} ultrasound tomography,'' in \emph{Medical Imaging 2022: Ultrasonic
  Imaging and Tomography}.\hskip 1em plus 0.5em minus 0.4em\relax SPIE, 2022,
  p. PC1203809.

\bibitem{gierach2022rapid}
G.~L. Gierach, M.~Sak, S.~Fan, R.~M. Pfeiffer, M.~Palakal, C.~Ramin,
  L.~Bey-Knight, M.~S. Simon, D.~Gorski, H.~Ali \emph{et~al.}, ``Rapid
  reductions in breast density following tamoxifen therapy as evaluated by
  whole-breast ultrasound tomography,'' \emph{Journal of Clinical Medicine},
  vol.~11, no.~3, p. 792, 2022.

\bibitem{kratkiewicz2022ultrasound}
K.~Kratkiewicz, A.~Pattyn, N.~Alijabbari, and M.~Mehrmohammadi, ``Ultrasound
  and photoacoustic imaging of breast cancer: Clinical systems, challenges, and
  future outlook,'' \emph{Journal of Clinical Medicine}, vol.~11, no.~5, p.
  1165, 2022.

\bibitem{littrup2021multicenter}
P.~J. Littrup, N.~Duric, M.~Sak, C.~Li, O.~Roy, R.~F. Brem, L.~H. Larsen, and
  M.~Yamashita, ``Multicenter study of whole breast stiffness imaging by
  ultrasound tomography ({S}oft{V}ue) for characterization of breast tissues
  and masses,'' \emph{Journal of Clinical Medicine}, vol.~10, no.~23, p. 5528,
  2021.

\bibitem{duric2020novel}
N.~Duric, P.~Littrup, M.~Sak, C.~Li, D.~Chen, O.~Roy, L.~Bey-Knight, and
  R.~Brem, ``A novel marker, based on ultrasound tomography, for monitoring
  early response to neoadjuvant chemotherapy,'' \emph{Journal of Breast
  Imaging}, vol.~2, no.~6, pp. 569--576, 2020.

\bibitem{jw2017}
J.~W. Wiskin, D.~T. Borup, E.~Iuanow, J.~Klock, and M.~W. Lenox, ``3-d
  nonlinear acoustic inverse scattering: Algorithm and quantitative results,''
  \emph{IEEE Transactions on Ultrasonics, Ferroelectrics, and Frequency
  Control}, vol.~64, no.~8, pp. 1161--1174, 2017.

\bibitem{wiskin2019quantitative}
J.~Wiskin, B.~Malik, R.~Natesan, and M.~Lenox, ``Quantitative assessment of
  breast density using transmission ultrasound tomography,'' \emph{Medical
  physics}, vol.~46, no.~6, pp. 2610--2620, 2019.

\bibitem{pratt2007sound}
R.~G. Pratt, L.~Huang, N.~Duric, and P.~Littrup, ``Sound-speed and attenuation
  imaging of breast tissue using waveform tomography of transmission ultrasound
  data,'' in \emph{Medical Imaging 2007: Physics of Medical Imaging}, vol.
  6510.\hskip 1em plus 0.5em minus 0.4em\relax International Society for Optics
  and Photonics, 2007, p. 65104S.

\bibitem{wang2015waveform}
K.~Wang, T.~Matthews, F.~Anis, C.~Li, N.~Duric, and M.~A. Anastasio, ``Waveform
  inversion with source encoding for breast sound speed reconstruction in
  ultrasound computed tomography,'' \emph{IEEE transactions on ultrasonics,
  ferroelectrics, and frequency control}, vol.~62, no.~3, pp. 475--493, 2015.

\bibitem{matthews2017regularized}
T.~P. Matthews, K.~Wang, C.~Li, N.~Duric, and M.~A. Anastasio, ``Regularized
  dual averaging image reconstruction for full-wave ultrasound computed
  tomography,'' \emph{IEEE transactions on ultrasonics, ferroelectrics, and
  frequency control}, vol.~64, no.~5, pp. 811--825, 2017.

\bibitem{duric2021potential}
N.~Duric, M.~Sak, and P.~J. Littrup, ``The potential role of the fat--glandular
  interface (fgi) in breast carcinogenesis: Results from an ultrasound
  tomography (ust) study,'' \emph{Journal of Clinical Medicine}, vol.~10,
  no.~23, p. 5615, 2021.

\bibitem{duric2020using}
N.~Duric, M.~Sak, S.~Fan, R.~M. Pfeiffer, P.~J. Littrup, M.~S. Simon, D.~H.
  Gorski, H.~Ali, K.~S. Purrington, R.~F. Brem \emph{et~al.}, ``Using whole
  breast ultrasound tomography to improve breast cancer risk assessment: A
  novel risk factor based on the quantitative tissue property of sound speed,''
  \emph{Journal of clinical medicine}, vol.~9, no.~2, p. 367, 2020.

\bibitem{bates2022probabilistic}
O.~Bates, L.~Guasch, G.~Strong, T.~C. Robins, O.~Calderon-Agudo, C.~Cueto,
  J.~Cudeiro, and M.~Tang, ``A probabilistic approach to tomography and adjoint
  state methods, with an application to full waveform inversion in medical
  ultrasound,'' \emph{Inverse Problems}, vol.~38, no.~4, p. 045008, 2022.

\bibitem{lucka2021high}
F.~Lucka, M.~P{\'e}rez-Liva, B.~E. Treeby, and B.~T. Cox, ``High resolution
  3{D} ultrasonic breast imaging by time-domain full waveform inversion,''
  \emph{Inverse Problems}, vol.~38, no.~2, p. 025008, 2021.

\bibitem{javaherian2021ray}
A.~Javaherian and B.~Cox, ``Ray-based inversion accounting for scattering for
  biomedical ultrasound tomography,'' \emph{Inverse Problems}, vol.~37, no.~11,
  p. 115003, 2021.

\bibitem{hormati2010robust}
A.~Hormati, I.~Jovanovi{\'c}, O.~Roy, and M.~Vetterli, ``Robust ultrasound
  travel-time tomography using the bent ray model,'' in \emph{Medical Imaging
  2010: Ultrasonic Imaging, Tomography, and Therapy}, vol. 7629.\hskip 1em plus
  0.5em minus 0.4em\relax International Society for Optics and Photonics, 2010,
  p. 76290I.

\bibitem{schreiman1984ultrasound}
J.~Schreiman, J.~Gisvold, J.~F. Greenleaf, and R.~Bahn, ``Ultrasound
  transmission computed tomography of the breast.'' \emph{Radiology}, vol. 150,
  no.~2, pp. 523--530, 1984.

\bibitem{carson1981breast}
P.~L. Carson, C.~R. Meyer, A.~L. Scherzinger, and T.~V. Oughton, ``Breast
  imaging in coronal planes with simultaneous pulse echo and transmission
  ultrasound,'' \emph{Science}, vol. 214, no. 4525, pp. 1141--1143, 1981.

\bibitem{andre1997high}
M.~P. Andr{\'e}, H.~S. Jan{\'e}e, P.~J. Martin, G.~P. Otto, B.~A. Spivey, and
  D.~A. Palmer, ``High-speed data acquisition in a diffraction tomography
  system employing large-scale toroidal arrays,'' \emph{International Journal
  of Imaging Systems and Technology}, vol.~8, no.~1, pp. 137--147, 1997.

\bibitem{duric2007detection}
N.~Duric, P.~Littrup, L.~Poulo, A.~Babkin, R.~Pevzner, E.~Holsapple, O.~Rama,
  and C.~Glide, ``Detection of breast cancer with ultrasound tomography:
  {F}irst results with the computed ultrasound risk evaluation ({CURE})
  prototype,'' \emph{Medical physics}, vol.~34, no.~2, pp. 773--785, 2007.

\bibitem{song2021design}
J.~Song, Q.~Zhang, L.~Zhou, Z.~Quan, S.~Wang, Z.~Liu, X.~Fang, Y.~Wu, Q.~Yang,
  H.~Yin \emph{et~al.}, ``Design and implementation of a modular and scalable
  research platform for ultrasound computed tomography,'' \emph{IEEE
  Transactions on Ultrasonics, Ferroelectrics, and Frequency Control}, 2021.

\bibitem{stotzka2003ultrasound}
R.~Stotzka, T.~O. M{\"u}ller, K.~Schlote-Holubek, and H.~Gemmeke, ``Ultrasound
  computer tomography for breast cancer diagnosis,'' in \emph{7th conference of
  the European Society for Engineering and Medicine}, 2003.

\bibitem{GemmekeBergerHopp2018_1000079770}
H.~Gemmeke, L.~Berger, T.~Hopp, M.~Zapf, W.~Tan, R.~Blanco, R.~Leys, I.~Peric,
  and N.~V. Ruiter, ``\BIBforeignlanguage{english}{The new generation of the
  {KIT} 3{D} {USCT}},'' in \emph{\BIBforeignlanguage{english}{Proceedings of
  the International Workshop on Medical Ultrasound Tomography: 1.- 3. Nov.
  2017, Speyer, Germany. Hrsg.: T. Hopp}}.\hskip 1em plus 0.5em minus
  0.4em\relax {KIT Scientific Publishing}, 2018, pp. 271--282, 54.02.02; LK 01.

\bibitem{malik2018quantitative}
B.~Malik, R.~Terry, J.~Wiskin, and M.~Lenox, ``Quantitative transmission
  ultrasound tomography: Imaging and performance characteristics,''
  \emph{Medical physics}, vol.~45, no.~7, pp. 3063--3075, 2018.

\bibitem{CUDEIROBLANCO20221995}
J.~Cudeiro-Blanco, C.~Cueto, O.~Bates, G.~Strong, T.~Robins, M.~Toulemonde,
  M.~Warner, M.-X. Tang, O.~C. Agudo, and L.~Guasch, ``Design and construction
  of a low-frequency ultrasound acquisition device for 2-d brain imaging using
  full-waveform inversion,'' \emph{Ultrasound in Medicine \& Biology}, vol.~48,
  no.~10, pp. 1995--2008, 2022.

\bibitem{9512083}
C.~Cueto, L.~Guasch, J.~Cudeiro, O.~C. Agudo, T.~Robins, O.~Bates, G.~Strong,
  and M.-X. Tang, ``Spatial response identification enables robust experimental
  ultrasound computed tomography,'' \emph{IEEE Transactions on Ultrasonics,
  Ferroelectrics, and Frequency Control}, vol.~69, no.~1, pp. 27--37, 2022.

\bibitem{duric2013breast}
N.~Duric, P.~Littrup, S.~Schmidt, C.~Li, O.~Roy, L.~Bey-Knight, R.~Janer,
  D.~Kunz, X.~Chen, J.~Goll \emph{et~al.}, ``Breast imaging with the
  {S}oft{V}ue imaging system: {F}irst results,'' in \emph{Medical Imaging 2013:
  Ultrasonic Imaging, Tomography, and Therapy}, vol. 8675.\hskip 1em plus 0.5em
  minus 0.4em\relax International Society for Optics and Photonics, 2013, p.
  86750K.

\bibitem{duric2014clinical}
N.~Duric, P.~Littrup, O.~Roy, C.~Li, S.~Schmidt, X.~Cheng, and R.~Janer,
  ``Clinical breast imaging with ultrasound tomography: {A} description of the
  {S}oft{V}ue system,'' \emph{The Journal of the Acoustical Society of
  America}, vol. 135, no.~4, pp. 2155--2155, 2014.

\bibitem{8197331}
J.~Song, S.~Wang, L.~Zhou, Y.~Peng, M.~Ding, and M.~Yuchi, ``A prototype system
  for ultrasound computer tomography with ring array,'' in \emph{2nd IET
  International Conference on Biomedical Image and Signal Processing (ICBISP
  2017)}, 2017, pp. 1--4.

\bibitem{9593866}
M.~Roberts, E.~Martin, M.~Brown, B.~Cox, and B.~Treeby, ``Transducer module
  development for an open-source ultrasound tomography system,'' in \emph{2021
  IEEE International Ultrasonics Symposium (IUS)}, 2021, pp. 1--4.

\bibitem{mervcep2019transmission}
E.~Mer{\v{c}}ep, J.~L. Herraiz, X.~L. De{\'a}n-Ben, and D.~Razansky,
  ``Transmission--reflection optoacoustic ultrasound (tropus) computed
  tomography of small animals,'' \emph{Light: Science \& Applications}, vol.~8,
  no.~1, p.~18, 2019.

\bibitem{jcm11051165}
K.~Kratkiewicz, A.~Pattyn, N.~Alijabbari, and M.~Mehrmohammadi, ``Ultrasound
  and photoacoustic imaging of breast cancer: Clinical systems, challenges, and
  future outlook,'' \emph{Journal of Clinical Medicine}, vol.~11, no.~5, 2022.

\bibitem{sandhu2015frequency}
G.~Sandhu, C.~Li, O.~Roy, S.~Schmidt, and N.~Duric, ``Frequency domain
  ultrasound waveform tomography: breast imaging using a ring transducer,''
  \emph{Physics in Medicine \& Biology}, vol.~60, no.~14, p. 5381, 2015.

\bibitem{li2022investigation}
F.~Li, U.~Villa, N.~Duric, and M.~A. Anastasio, ``Investigation of an
  elevation-focused transducer model for three-dimensional full-waveform
  inversion in ultrasound computed tomography,'' in \emph{Medical Imaging 2022:
  Ultrasonic Imaging and Tomography}, vol. 12038.\hskip 1em plus 0.5em minus
  0.4em\relax SPIE, 2022, pp. 206--214.

\bibitem{poudel2019compensation}
J.~Poudel, L.~A. Forte, and M.~A. Anastasio, ``Compensation of 3{D}-2{D} model
  mismatch in ultrasound computed tomography with the aid of convolutional
  neural networks (conference presentation),'' in \emph{Medical Imaging 2019:
  Ultrasonic Imaging and Tomography}, vol. 10955.\hskip 1em plus 0.5em minus
  0.4em\relax International Society for Optics and Photonics, 2019, p. 1095507.

\bibitem{martin2016simulating}
E.~Martin, Y.~T. Ling, and B.~E. Treeby, ``Simulating focused ultrasound
  transducers using discrete sources on regular cartesian grids,'' \emph{IEEE
  transactions on ultrasonics, ferroelectrics, and frequency control}, vol.~63,
  no.~10, pp. 1535--1542, 2016.

\bibitem{wise2019representing}
E.~S. Wise, B.~Cox, J.~Jaros, and B.~E. Treeby, ``Representing arbitrary
  acoustic source and sensor distributions in {F}ourier collocation methods,''
  \emph{The Journal of the Acoustical Society of America}, vol. 146, no.~1, pp.
  278--288, 2019.

\bibitem{li2021NBPs3D}
\BIBentryALTinterwordspacing
F.~Li, U.~Villa, and M.~Anastasio, ``{3D Numerical Breast Phantoms and
  Ring-Array USCT measurements (3 rings)},'' 2023. [Online]. Available:
  \url{https://doi.org/10.7910/DVN/8JVLAE}
\BIBentrySTDinterwordspacing

\bibitem{lions2012non}
J.~L. Lions and E.~Magenes, \emph{Non-homogeneous boundary value problems and
  applications: Vol. 1}.\hskip 1em plus 0.5em minus 0.4em\relax Springer
  Science \& Business Media, 2012, vol. 181.

\bibitem{evans2012introduction}
L.~C. Evans, \emph{An introduction to stochastic differential equations}.\hskip
  1em plus 0.5em minus 0.4em\relax American Mathematical Soc., 2012, vol.~82.

\bibitem{treeby2010modeling}
B.~E. Treeby and B.~T. Cox, ``Modeling power law absorption and dispersion for
  acoustic propagation using the fractional laplacian,'' \emph{The Journal of
  the Acoustical Society of America}, vol. 127, no.~5, pp. 2741--2748, 2010.

\bibitem{chen2003modified}
W.~Chen and S.~Holm, ``Modified szabo’s wave equation models for lossy media
  obeying frequency power law,'' \emph{The Journal of the Acoustical Society of
  America}, vol. 114, no.~5, pp. 2570--2574, 2003.

\bibitem{szabo1995causal}
T.~L. Szabo, ``Causal theories and data for acoustic attenuation obeying a
  frequency power law,'' \emph{The Journal of the Acoustical Society of
  America}, vol.~97, no.~1, pp. 14--24, 1995.

\bibitem{1503968}
K.~Waters, J.~Mobley, and J.~Miller, ``Causality-imposed (kramers-kronig)
  relationships between attenuation and dispersion,'' \emph{IEEE Transactions
  on Ultrasonics, Ferroelectrics, and Frequency Control}, vol.~52, no.~5, pp.
  822--823, 2005.

\bibitem{zhang2012efficient}
Z.~Zhang, L.~Huang, and Y.~Lin, ``Efficient implementation of ultrasound
  waveform tomography using source encoding,'' in \emph{Medical Imaging 2012:
  Ultrasonic Imaging, Tomography, and Therapy}, vol. 8320.\hskip 1em plus 0.5em
  minus 0.4em\relax SPIE, 2012, pp. 22--31.

\bibitem{moghaddam2013new}
P.~P. Moghaddam, H.~Keers, F.~J. Herrmann, and W.~A. Mulder, ``A new
  optimization approach for source-encoding full-waveform inversion,''
  \emph{Geophysics}, vol.~78, no.~3, pp. R125--R132, 2013.

\bibitem{krebs2009fast}
J.~R. Krebs, J.~E. Anderson, D.~Hinkley, R.~Neelamani, S.~Lee, A.~Baumstein,
  and M.-D. Lacasse, ``Fast full-wavefield seismic inversion using encoded
  sources,'' \emph{Geophysics}, vol.~74, no.~6, pp. WCC177--WCC188, 2009.

\bibitem{smith1978real}
S.~Smith, O.~Von~Ramm, J.~Kisslo, and F.~Thurstone, ``Real time ultrasound
  tomography of the adult brain.'' \emph{Stroke}, vol.~9, no.~2, pp. 117--122,
  1978.

\bibitem{song2013liquid}
C.~Song, L.~Xi, and H.~Jiang, ``Liquid acoustic lens for photoacoustic
  tomography,'' \emph{Optics letters}, vol.~38, no.~15, pp. 2930--2933, 2013.

\bibitem{o1949theory}
H.~O'Neil, ``Theory of focusing radiators,'' \emph{The Journal of the
  Acoustical Society of America}, vol.~21, no.~5, pp. 516--526, 1949.

\bibitem{hansen2001fundamentals}
C.~H. Hansen, ``Fundamentals of acoustics,'' \emph{Occupational Exposure to
  Noise: Evaluation, Prevention and Control. World Health Organization},
  vol.~1, no.~3, pp. 23--52, 2001.

\bibitem{Guyomarfocu}
\BIBentryALTinterwordspacing
D.~Guyomar and J.~Powers, ``{Transient fields radiated by curved
  surfaces—Application to focusing},'' \emph{The Journal of the Acoustical
  Society of America}, vol.~76, no.~5, pp. 1564--1572, 11 1984. [Online].
  Available: \url{https://doi.org/10.1121/1.391467}
\BIBentrySTDinterwordspacing

\bibitem{Penttinen_1976}
\BIBentryALTinterwordspacing
A.~Penttinen and M.~Luukkala, ``Sound pressure near the focal area of an
  ultrasonic lens,'' \emph{Journal of Physics D: Applied Physics}, vol.~9,
  no.~13, p. 1927, sep 1976. [Online]. Available:
  \url{https://dx.doi.org/10.1088/0022-3727/9/13/013}
\BIBentrySTDinterwordspacing

\bibitem{4154642}
X.-h. Yan, Y.-p. Zhang, K.-h. Liu, and Y.~Liu, ``Numerical calculation of the
  sound field focused by acoustic lens with an arbitrary axisymmetric sound
  speed distribution,'' \emph{IEEE Transactions on Ultrasonics, Ferroelectrics,
  and Frequency Control}, vol.~54, no.~4, pp. 823--829, 2007.

\bibitem{wu2002theoretical}
X.~Wu and M.~Sherar, ``Theoretical evaluation of moderately focused spherical
  transducers and multi-focus acoustic lens/transducer systems for ultrasound
  thermal therapy,'' \emph{Physics in Medicine \& Biology}, vol.~47, no.~9, p.
  1603, 2002.

\bibitem{911727}
Y.~Huo and Y.~Chen, ``Simulation of field characteristics of the focused
  axisymmetrically curved surface transducers,'' \emph{IEEE Transactions on
  Ultrasonics, Ferroelectrics, and Frequency Control}, vol.~48, no.~2, pp.
  445--451, 2001.

\bibitem{marechal2007lens}
P.~Mar{\'e}chal, F.~Levassort, L.-P. Tran-Huu-Hue, and M.~Lethiecq,
  ``Lens-focused transducer modeling using an extended klm model,''
  \emph{Ultrasonics}, vol.~46, no.~2, pp. 155--167, 2007.

\bibitem{kostli2003two}
K.~P. K{\"o}stli and P.~C. Beard, ``Two-dimensional photoacoustic imaging by
  use of {F}ourier-transform image reconstruction and a detector with an
  anisotropic response,'' \emph{Applied optics}, vol.~42, no.~10, pp.
  1899--1908, 2003.

\bibitem{ding2017efficient}
L.~Ding, X.~L. De{\'a}n-Ben, and D.~Razansky, ``Efficient 3-d model-based
  reconstruction scheme for arbitrary optoacoustic acquisition geometries,''
  \emph{IEEE transactions on medical imaging}, vol.~36, no.~9, pp. 1858--1867,
  2017.

\bibitem{5560859}
K.~Wang, S.~A. Ermilov, R.~Su, H.-P. Brecht, A.~A. Oraevsky, and M.~A.
  Anastasio, ``An imaging model incorporating ultrasonic transducer properties
  for three-dimensional optoacoustic tomography,'' \emph{IEEE Transactions on
  Medical Imaging}, vol.~30, no.~2, pp. 203--214, 2011.

\bibitem{verweij2014simulation}
M.~Verweij, B.~Treeby, K.~Van~Dongen, and L.~Demi, ``Simulation of ultrasound
  fields,'' \emph{Comprehensive biomedical physics}, pp. 465--499, 2014.

\bibitem{treeby2010k}
B.~E. Treeby and B.~T. Cox, ``k-{W}ave: {MATLAB} toolbox for the simulation and
  reconstruction of photoacoustic wave fields,'' \emph{Journal of biomedical
  optics}, vol.~15, no.~2, p. 021314, 2010.

\bibitem{norton1999iterative}
S.~J. Norton, ``Iterative inverse scattering algorithms: Methods of computing
  frechet derivatives,'' \emph{The Journal of the Acoustical Society of
  America}, vol. 106, no.~5, pp. 2653--2660, 1999.

\bibitem{plessix2006review}
R.-E. Plessix, ``A review of the adjoint-state method for computing the
  gradient of a functional with geophysical applications,'' \emph{Geophysical
  Journal International}, vol. 167, no.~2, pp. 495--503, 2006.

\bibitem{perez2017time}
M.~Pérez-Liva, J.~L. Herraiz, J.~M. Udías, E.~Miller, B.~T. Cox, and B.~E.
  Treeby, ``Time domain reconstruction of sound speed and attenuation in
  ultrasound computed tomography using full wave inversion,'' \emph{The Journal
  of the Acoustical Society of America}, vol. 141, no.~3, pp. 1595--1604, 2017.

\bibitem{5610565}
T.~Kundu, D.~Placko, E.~K. Rahani, T.~Yanagita, and C.~M. Dao, ``Ultrasonic
  field modeling: a comparison of analytical, semi-analytical, and numerical
  techniques,'' \emph{IEEE Transactions on Ultrasonics, Ferroelectrics, and
  Frequency Control}, vol.~57, no.~12, pp. 2795--2807, 2010.

\bibitem{treeby2012modeling}
B.~E. Treeby, J.~Jaros, A.~P. Rendell, and B.~Cox, ``Modeling nonlinear
  ultrasound propagation in heterogeneous media with power law absorption using
  ak-space pseudospectral method,'' \emph{The Journal of the Acoustical Society
  of America}, vol. 131, no.~6, pp. 4324--4336, 2012.

\bibitem{samarasinghe2017acoustic}
P.~Samarasinghe, T.~D. Abhayapala, and W.~Kellermann, ``Acoustic reciprocity:
  An extension to spherical harmonics domain,'' \emph{The Journal of the
  Acoustical Society of America}, vol. 142, no.~4, pp. EL337--EL343, 2017.

\bibitem{9537158}
F.~Li, U.~Villa, S.~Park, and M.~A. Anastasio, ``3-d stochastic numerical
  breast phantoms for enabling virtual imaging trials of ultrasound computed
  tomography,'' \emph{IEEE Transactions on Ultrasonics, Ferroelectrics, and
  Frequency Control}, vol.~69, no.~1, pp. 135--146, 2022.

\bibitem{Multiparametersinv}
\BIBentryALTinterwordspacing
U.~Taskin and K.~W.~A. van Dongen, ``{Multi-parameter inversion with the aid of
  particle velocity field reconstruction},'' \emph{The Journal of the
  Acoustical Society of America}, vol. 147, no.~6, pp. 4032--4040, 06 2020.
  [Online]. Available: \url{https://doi.org/10.1121/10.0001396}
\BIBentrySTDinterwordspacing

\bibitem{li3dfwi}
F.~Li, U.~Villa, N.~Duric, and M.~A. Anastasio, ``{3D full-waveform inversion
  in ultrasound computed tomography employing a ring-array},'' in \emph{Medical
  Imaging 2023: Ultrasonic Imaging and Tomography}, vol. 12470, International
  Society for Optics and Photonics.\hskip 1em plus 0.5em minus 0.4em\relax
  SPIE, 2023, p. 124700K.

\bibitem{yuan2023full}
Y.~Yuan, Y.~Zhao, N.~Zhang, Y.~Xiao, J.~Jin, N.~Feng, and Y.~Shen,
  ``Full-waveform inversion for breast ultrasound tomography using line-shape
  modeled elements,'' \emph{Ultrasound in Medicine \& Biology}, vol.~49, no.~5,
  pp. 1070--1081, 2023.

\bibitem{cottrell2009isogeometric}
J.~A. Cottrell, T.~J. Hughes, and Y.~Bazilevs, \emph{Isogeometric analysis:
  toward integration of CAD and FEA}.\hskip 1em plus 0.5em minus 0.4em\relax
  John Wiley \& Sons, 2009.

\bibitem{komatitsch2018specfem}
D.~Komatitsch, J.~Vilotte, J.~Tromp, and development team, ``{SPECFEM} 3{D}
  {Cartesian} user manual version 3.0,'' \emph{CNRS (France), Princeton
  University (USA) i ETH Z{\"u}rich (Switzerland)}, 2018.

\bibitem{schoeder2019exwave}
S.~Schoeder, W.~A. Wall, and M.~Kronbichler, ``Exwave: A high performance
  discontinuous galerkin solver for the acoustic wave equation,''
  \emph{SoftwareX}, vol.~9, pp. 49--54, 2019.

\bibitem{roy2011robust}
O.~Roy, I.~Jovanovi{\'c}, N.~Duric, L.~Poulo, and M.~Vetterli, ``Robust array
  calibration using time delays with application to ultrasound tomography,'' in
  \emph{Medical Imaging 2011: Ultrasonic Imaging, Tomography, and Therapy},
  vol. 7968.\hskip 1em plus 0.5em minus 0.4em\relax SPIE, 2011, pp. 46--56.

\bibitem{dukhin2010fundamentals}
A.~S. Dukhin and P.~J. Goetz, ``Fundamentals of acoustics in homogeneous
  liquids: Longitudinal rheology,'' in \emph{Studies in interface
  science}.\hskip 1em plus 0.5em minus 0.4em\relax Elsevier, 2010, vol.~24, pp.
  91--125.

\bibitem{2das}
\BIBentryALTinterwordspacing
B.~E. Treeby, E.~S. Wise, F.~Kuklis, J.~Jaros, and B.~T. Cox, ``{Nonlinear
  ultrasound simulation in an axisymmetric coordinate system using a k-space
  pseudospectral methoda)},'' \emph{The Journal of the Acoustical Society of
  America}, vol. 148, no.~4, pp. 2288--2300, 10 2020. [Online]. Available:
  \url{https://doi.org/10.1121/10.0002177}
\BIBentrySTDinterwordspacing

\bibitem{berenger1994perfectly}
J.-P. Berenger, ``A perfectly matched layer for the absorption of
  electromagnetic waves,'' \emph{Journal of computational physics}, vol. 114,
  no.~2, pp. 185--200, 1994.

\end{thebibliography}
\bibliographystyle{IEEEtran}

\end{document}